\documentclass[12pt,numbers,sort&compress]{article}

\usepackage{latexsym,amsmath,amssymb,theorem,epsfig} 
\usepackage{natbib}
\usepackage{graphicx}
\usepackage{color}

\usepackage[isu,bf]{caption2}
\numberwithin{equation}{section}

\newcommand{\ns}{\normalsize}

\DeclareMathOperator{\Tr}{Tr}
\DeclareMathOperator{\RE}{Re}
\DeclareMathOperator{\IM}{Im}
\DeclareMathOperator{\Pfaff}{Pfaff}
\DeclareMathOperator{\Hom}{Hom}
\newcommand{\Z}{\mathbb{Z}}
\newcommand{\ZZZ}{{\ensuremath{\Z_3\times\Z_3}}}
\newcommand{\B}[1]{\ensuremath{B_{#1}}}

\newcommand{\Xt}{{\ensuremath{\widetilde{X}}}}
\newcommand{\oB}[1]{\ensuremath{\mathcal{O}_{\B{#1}}}}
\newcommand{\oXt}{\ensuremath{\mathcal{O}_{\Xt}}}
\newcommand{\V}[1]{\ensuremath{\mathcal{V}_{#1}}}
\newcommand{\Vt}{{\ensuremath{\widetilde{\V{}}}}}
\newcommand{\C}{\mathbb{C}}
\newcommand{\CP}[1]{\mathbb{P}^{#1}}
\newcommand{\IP}[1]{\CP{#1}}
\newcommand{\Kcone}{\ensuremath{\mathcal{K}}}
\newtheorem{proposition}{Proposition}

\theoremstyle{plain} 

{\theorembodyfont{\rmfamily} }

\begin{document}

%%%%%%%%%%%%%%%%%%%%%%%%%%%%%%%%%%%%%%%%%%%%%%%%%%%%%%%%%%%%%%%%%%%%%%

\begin{titlepage}

\vspace{-5cm}

\title{
   \hfill{\ns hep-th/0603088} \\[1em]
%\\[-4mm]
%   \hfill{\ns UPR-} \\[1em]
   {\LARGE Stabilizing Moduli \\
     with a Positive Cosmological Constant \\
     in Heterotic M-Theory} 
\\[1em] } 
\author{
   Volker Braun
   and Burt A. Ovrut \\[0.5em]
   {\ns Department of Physics, University of Pennsylvania} \\[-0.4em]
   {\ns Philadelphia, PA 19104--6396}\\}
      
\date{}

\maketitle
\vspace{1cm}

\begin{abstract}

  It is shown that strongly coupled heterotic $M$-theory with
  anti-five-branes in the $S^{1}/{\mathbb{Z}}_{2}$ bulk space can have
  meta-stable vacua which break $N=1$ supersymmetry and have a small,
  positive cosmological constant. This is demonstrated for the
  ``minimal'' heterotic standard model. This vacuum has the exact MSSM
  matter spectrum in the observable sector, a trivial hidden sector
  vector bundle and both five-branes and anti-five-branes in the bulk
  space. The K\"ahler moduli for which the cosmological constant has
  phenomenologically acceptable values are shown to also render the
  observable sector vector bundle slope-stable. A corollary of this
  result is that strongly coupled $M$-theory vacua with only
  five-branes in the $S^{1}/{\mathbb{Z}}_{2}$ interval may have
  stabilized moduli, but at a supersymmetry preserving minimum with a
  large, negative cosmological constant.

\end{abstract}

\thispagestyle{empty}

\end{titlepage}

%%%%%%%%%%%%%%%%%%%%%%%%%%%%%%%%%%%%%%%%%%%%%%%%%%%%%%%%%%%%%%%%%%%%%%%

\section{Introduction}
\label{sec:intro}

In a series of papers, vacuum states of the $E_{8} \times E_{8}$
heterotic superstring were presented whose observable sectors have the
matter spectrum of the minimal supersymmetric standard model (MSSM)
with the addition of one extra pair of Higgs-Higgs conjugate
fields~\cite{Braun:2005ux, Braun:2005bw, Braun:2005zv}. Subsequently,
it was shown~\cite{Braun:2005nv, Ovrut:2006yr} that a subclass of
these theories have \emph{exactly} the matter content of the MSSM.
Since the observable sector matter spectra are realistic, these ground
states are called ``heterotic standard models''.  The vacuum with
exactly the MSSM matter spectrum is called the ``minimal'' heterotic
standard model.  In the last year similar observable sectors were
found using orbifolds~\cite{Buchmuller:2005jr} or geometrically with a
supersymmetric hidden sector~\cite{Bouchard:2005ag, Bouchard:2006dn}.
Other approaches can be found in~\cite{Greene:1986bm, Greene:1986jb,
  Greene:1986ar, Blumenhagen:2005zh, Blumenhagen:2006ux,
  Andreas:2004ja}

The vacua~\cite{Braun:2005ux, Braun:2005bw, Braun:2005zv,
  Braun:2005nv} are constructed by compactifying the $E_{8} \times
E_{8}$ heterotic string on a torus-fibered Calabi-Yau threefold whose
fundamental group is ${\mathbb Z}_{3} \times {\mathbb
  Z}_{3}$~\cite{Braun:2004xv}. The observable sector contains a
holomorphic vector bundle $V$ with structure group $SU(4)$. This
bundle was proven to be slope-stable for the two Higgs pair vacua
in~\cite{Gomez:2005ii} and for the minimal heterotic standard model
in~\cite{Braun:2006ae}.  Hence, in all cases the observable sector
admits a gauge connection satisfying the hermitian Yang-Mills
equations. The $SU(4)$ structure group breaks $E_{8}$ down to
$Spin(10)$. The discovery of non-vanishing neutrino masses requires
that realistic supersymmetric theories contain right-handed
neutrinos~\cite{Langacker:2004xy, Giedt:2005vx}.  It is well-known
that the $\mathbf{16}$ representation of $Spin(10)$ is composed of an
entire family of quarks/leptons, including a right-handed neutrino.
For this reason, an $SU(4)$ vector bundle was chosen for the
observable sector of heterotic standard models.  A formalism for
computing the number of vector bundle moduli was presented
in~\cite{Braun:2005fk} and applied to the observable sector bundles.
For example, the number of vector bundle moduli in the observable
sector of the minimal heterotic standard model is thirteen. The
$Spin(10)$ gauge group is then broken by ${\mathbb Z}_{3} \times
{\mathbb Z}_{3}$ Wilson lines. Since ${\mathbb Z}_{3} \times {\mathbb
  Z}_{3}$ is Abelian, the low energy theory consists of the standard
model gauge group, $SU(3)_{C} \times SU(2)_{L} \times U(1)_{Y}$, times
an additional gauge group, $U(1)_{B-L}$, whose charges are the $B-L$
quantum numbers.

In addition to having realistic spectra, the observable sector of
heterotic standard models also satisfies important phenomenological
constraints. Consider nucleon decay in this context.  The additional
$U(1)_{B-L}$ symmetry, if spontaneously broken at a low mass scale,
suppresses $\Delta L=1$ and $\Delta B=1$ dimension four operators in
the effective theory. These vacua exhibit natural doublet-triplet
splitting~\cite{Witten:2001bf, Donagi:2004su, Braun:2005ux}.  This
eliminates color triplet induced dimension five operators which can
lead to rapid nucleon decay. The unification scale of heterotic
standard models is of ${\cal{O}}(10^{16}GeV)$. Hence, nucleon decay
via heavy $Spin(10)$ vector bosons is sufficiently suppressed. Taken
together, we see that heterotic standard models naturally reduce the
nucleon decay rate to a level consistent with experimental
bounds~\cite{Nath:2006ut, Tatar:2006dc}. Formalisms for computing the
Higgs $\mu$-terms and Yukawa couplings in the observable sectors of
heterotic standard models were presented in~\cite{Braun:2005xp}
and~\cite{Braun:2006me} respectively.  In the minimal theory, for
example, it was shown that the cubic moduli-Higgs-Higgs conjugate
terms in the superpotential vanish due to ``geometric'' effects.
Therefore, non-vanishing Higgs $\mu$-terms only arise from higher
order interactions and, hence, are naturally
suppressed~\cite{Giudice:1988yz, Weinberg:2000cr}. Minimal heterotic
standard models were also shown to have an interesting texture of
Yukawa couplings which renders the first quark/lepton family naturally
light.

The hidden sector of heterotic standard models can be constructed in
two ways. Both must have a slope-stable holomorphic vector bundle on
the hidden orbifold plane.  In the first approach, one allows only
five-branes in the bulk space. The hidden sector bundle and the
cohomology class of the five-brane are then chosen so as to saturate
the anomaly cancellation condition. For heterotic standard models with
five-branes, it was shown in~\cite{Braun:2006ae} that the Chern class
of the hidden sector bundle satisfies a strong necessary condition,
the Bogomolov bound, for slope-stability. However, explicit
slope-stable hidden sector bundles were not constructed
in~\cite{Braun:2005nv, Braun:2006ae}. Note that vacua of this type are
$N=1$ supersymmetric. The only obvious way to break supersymmetry in
this context is via gaugino condensation in the hidden sector.
However, as was shown in a simplified theory in~\cite{Lukas:1997rb}
and reviewed and extended to a larger context in this paper, gaugino
condensation alone is not sufficient to break supersymmetry in
strongly coupled heterotic vacua. Furthermore, the minimum of the
moduli potential energy will have a large, negative cosmological
constant, consistent with the preservation of supersymmetry.  Given
the fact that supersymmetry is broken at energies below the
electroweak scale, and the observed very small, positive cosmological
constant~\cite{Riess:1998cb}, hidden sectors of this type would not
appear to be of phenomenological interest.

The second approach, which in many ways is mathematically simpler, is
to allow \emph{both} five-branes and anti-five-branes in the bulk
space. The hidden sector bundle and the five-brane/anti-five-brane
curves are then chosen to satisfy the anomaly cancellation condition.
The simplest way to do this is to take the hidden sector vector bundle
to be trivial, which is trivially slope-stable. If one assumes that 
there are no Wilson lines in the hidden sector then the
hidden sector gauge group is $E_{8}$.  Given a specific Calabi-Yau
threefold and observable sector bundle, the classes of the
five-brane/anti-five-brane curves are explicitly fixed by the anomaly
condition. The appearance of an anti-five-brane in the bulk space can
potentially solve both problems inherent in the first approach. First
of all, it explicitly breaks supersymmetry and, hence, one expects
supersymmetry breaking operators in the effective theory. Secondly, as
with the anti D-branes of the Type IIB theories discussed
in~\cite{Kachru:2003aw, Kachru:2002gs}, anti-five-branes give a
positive contribution to the effective theory. This allows the minimum
of the moduli potential energy function to be ``uplifted'' to a small,
positive cosmological constant.  Hence, the second approach to the
hidden sector in heterotic standard models would appear to be more
suitable to construct realistic models of particle physics and
cosmology.

In this paper we will discuss the structure of heterotic standard
models where the hidden sector has a trivial bundle with no Wilson
lines and there are both five-branes and anti-five-branes in the bulk
space. Specifically, we will do the following. In
Section~\ref{sec:action}, we review the structure of heterotic
$M$-theory vacua and present the most general K\"ahler potentials and
superpotentials in this context.  In Section~\ref{sec:modfix} it is
shown that in strongly coupled heterotic vacua with only five-branes
in the bulk space and gaugino condensation in the hidden sector, the
values of all moduli can be fixed. However, the minimum of the
potential energy has a large, negative cosmological constant and does
not break supersymmetry. This is proven within a slightly simplified
context.  Specifically, we consider the dilaton, all K\"ahler and
complex structure moduli and the translation modulus of the bulk
five-brane. However, to make the analysis tractable only a single
vector bundle modulus is assumed. We also introduce string instantons
on representative curves only. A further analysis of string instantons
is, at present, impossible since the complete instanton series is
unknown.  Within this context,in Section~\ref{sec:anti}, we show that
the addition of an anti-five-brane to the bulk space and choosing the
hidden sector bundle to be trivial continues to admit a vacuum which
stabilizes all moduli. However, by appropriately choosing the K\"ahler
moduli one can make the minimum of the potential energy have a small,
positive cosmological constant consistent with the observed value.
This minimum will also break $N=1$ supersymmetry. Finally, in
Section~\ref{sec:dS} we apply this formalism to the moduli of the
minimal heterotic standard model~\cite{Braun:2005nv}. All moduli are
stabilized in this context. Furthermore, the minimum of the potential
can have a phenomenologically acceptable positive cosmological
constant and break supersymmetry.  This occurs for K\"ahler moduli for
which the observable sector vector bundle is slope-stable.

Finally, we want to point out that some of the results derived in this
paper were anticipated, in a simpler context,
in~\cite{Buchbinder:2003pi, Buchbinder:2004im}.

\section{$\mathbf{E_{8} \times E_{8}}$ Heterotic Vacua with Five-Branes}
\label{sec:action}

\subsection*{Basic Structure of the Vacuum}

In this section, we will consider $E_{8} \times E_{8}$ strongly coupled
heterotic string theory compactified on the space
\begin{equation}
  M=\mathbb{R}^4 \times X \times S^{1}/Z_2,
  \label{1}
\end{equation}
where $X$ is a Calabi-Yau threefold and $S^{1}/Z_{2}$ is an interval in
the eleventh-dimension. Furthermore, we will choose the Calabi-Yau 
threefold $X$ to be elliptically- or torus-fibered over either an Enriques 
surface $\cal{E}$, a del Pezzo surface $d{\mathbb P}_{i}$, $i=1,\dots,9$, or
a Hirzebruch surface ${\mathbb F}_r$, for integers $r \geq 0$. The threefold
may have either trivial or non-trivial fundamental group. 

Denote by $v_{CY}$ and $\pi \rho$ the reference volume of the Calabi-Yau
threefold and the reference length of the interval in the eleventh dimension
respectively. The physical volume and length are obtained by multiplying
them by the appropriate moduli. To achieve the correct phenomenological 
values for the four-dimensional Newton and gauge coupling parameters,
\begin{equation}
  M_{Pl} \sim 10^{19}GeV, \quad \alpha_{GUT} \sim \frac{1}{25},
  \label{2}
\end{equation}
we choose the inverse reference radius of the Calabi-Yau threefold 
and the inverse reference length of the eleventh dimension to be
\begin{equation}
  v_{CY}^{-1/6} \sim 10^{16}GeV, \quad (\pi \rho)^{-1} \sim 10^{14}GeV
  \label{3}
\end{equation}
respectively.

Let us list the complex moduli fields arising from such a compactification. 
They are the dilaton $S$, the $h^{1,1}$ moduli $T^{I}$ and the
$h^{2,1}$ moduli $Z_{\alpha}$. These moduli will be taken to be dimensionless.
Note that
\begin{equation}
  \RE S=V, \quad \RE T^{I}= R a^{I} V^{-1/3},
  \label{3a}
\end{equation}
where $V$ is the volume modulus of the Calabi-Yau threefold, R is the
modulus associated with the length of the $S^{1}/Z_2$ interval and the
$a^{I}$ satisfy the constraint
\begin{equation}
  V=\frac{1}{6}d_{IJK}a^{I}a^{J}a^{K}
  \label{3b}
\end{equation}
with $d_{IJK}$ the Calabi-Yau intersection numbers.  Given the
reference lengths chosen in~\eqref{3}, the dimensionless moduli $\RE
S$ and $R$ must be stabilized at the values
\begin{equation}
  \RE S \sim 1, \quad R \sim 1.
  \label{4}
\end{equation}

In addition to the geometrical compactification manifold~\eqref{1},
one must specify a slope-stable holomorphic vector bundle on the
orbifold plane at each end of the interval. On the observable brane,
one chooses a vector bundle $V$ with structure group $G \subseteq
E_{8}$ such that the low energy theory is supersymmetric and
realistic. For example, bundles leading to the exact MSSM
spectrum~\cite{Braun:2005nv, Bouchard:2005ag}, extensions of the
MSSM~\cite{Donagi:2000zf, Donagi:2004ia, Donagi:2004qk, Donagi:2004ub,
  Braun:2005ux, Braun:2005bw, Braun:2005zv} and $GUT$
theories~\cite{Greene:1986bm, Diaconescu:2005pc} at low energy have
been constructed. These vector bundles give rise to complex vector
bundle moduli $\phi_{a}$. The hidden orbifold plane must also support
a slope-stable holomorphic vector bundle $V'$ with structure group $G'
\subseteq E_{8}$. In this paper, we will always assume that $V'$ is
the trivial bundle. This bundle is trivially slope-stable and leads to
a low energy hidden sector gauge group of $E_{8}$ with no additional
moduli.

Finally, we allow for the possibility that there are five-branes and
anti-five-branes in the bulk space between the orbifold planes.  Let
$[\cal{W}]$ and $[\bar{\cal{W}}]$ be the Poincare dual of the curves
on which these branes and anti-branes respectively are wrapped. Then
the cancellation of quantum anomalies requires that
\begin{equation}
  c_{2}(V)-c_{2}(TX)+[{\cal{W}}]-[\bar{\cal{W}}]=0.
  \label{5}
\end{equation}
Note that $c_{2}(V')=0$ since $V'$ has been chosen to be trivial.
Generically, the curve class satisfying~\eqref{5} is neither effective
nor anti-effective, corresponding to both wrapped five-branes and
anti-five-branes.  However, in this section, we will assume that $TX$
and $V$ are such that the curve is strictly effective, leading to
wrapped five-branes only.  Later in the paper we will loosen this
assumption and allow for anti-branes as well. We will further assume
that the curve is chosen so as to correspond to the wrapping of a
single five-brane. In this case, the five-brane contributes another
complex modulus $\bf Y \rm$ to the low energy
theory~\cite{Derendinger:2000gy, Lima:2001jc, Lima:2001nh}, where
\begin{equation}
  \RE {\bf Y \rm}=\frac{Y \RE{\cal{T}}}{\pi\rho}
  \label{5a}
\end{equation}
and Y is the position of the five-brane in the eleventh dimension. The
superfield ${\cal{T}}$ is defined as follows.  Let $\omega_{I}$,
$I=1,\dots,h^{1,1}$ be a basis for $H^{1,1}$ and $z^{J}$,
$J=1,\dots,h^{1,1}$ be the dual basis of $H_{2}$ where
\begin{equation}
  \frac{1}{v_{CY}^{1/3}}\int_{z^{J}}{\omega_{I}}=\delta^{I}_{J}.
  \label{5b}
\end{equation}
The curve $z_{5}$ on which the five-brane is wrapped can be expanded
as $z_{5}=c_{I}z^{I}$. Then ${\cal{T}}$ is defined to be
\begin{equation}
  {\cal{T}}=c_{I}T^{I}.
  \label{5c}
\end{equation}
It is straightforward to show that
\begin{equation}
  \RE {\cal{T}} =\frac{1}{v_{CY}^{1/3}}\int_{z_{5}}{\omega_{T}},
  \label{5d}
\end{equation}
where 
\begin{equation}
  \omega_{T}=\RE T^{I}\omega_{I}.
  \label{5e}
\end{equation}

\subsection*{K\"ahler Potentials}

The K\"ahler potential for the $S$ and $T^{I}$ moduli was computed
in~\cite{Lukas:1997fg} and is given by
\begin{equation}
  K_{S,T}= -M_{Pl}^{2}\ln(S+\bar{S})-M_{Pl}^{2}\ln\big(\frac{1}{6}d_{IJK}(T+\bar{T})^{I}(T+\bar{T})^{J}
  (T+\bar{T})^{K} \big).
  \label{6}
\end{equation}
The K\"ahler potential for the complex structure moduli $Z_{\alpha}$ was
found in~\cite{Candelas:1990pi} to be
\begin{equation}
  K_{Z}=-M_{Pl}^{2}\ln(-i\int_{X}{\Omega \wedge \bar{\Omega}}),
  \label{7}
\end{equation}
where $\Omega$ is the holomorphic $(3,0)$ form.
In~\cite{Lukas:1998hk}, the K\"ahler potential for the five-brane
modulus $\bf Y \rm$ was calculated.  It was shown that
\begin{equation}
  K_{5}=2M_{Pl}^{2}\tau_{5} \frac{(\bf Y \rm + \bar{\bf Y \rm})^{2}}
  {(S+\bar{S})({\cal{T}}+\bar{\cal{T}})}
  \,,
  \label{10}
\end{equation}
with $\tau_{5}$ and $T_{5}$ given by 
\begin{align}
  \tau_{5} ~&= \frac{T_{5}v_{CY}^{1/3}(\pi\rho)^{2}}{M_{Pl}^{2}}
  \,,
  \label{11}
  \\
  T_{5} ~&= (2\pi)^{1/3}\frac{1}{(2 \kappa_{11}^{2})^{2/3}}
  \,.
  \label{12}
\end{align}
The eleven-dimensional gravitational coupling parameter $\kappa_{11}$
is related to the four-dimensional Planck mass as
\begin{equation}
  \kappa_{11}^{2}=\frac{\pi \rho v_{CY}}{M_{Pl}^{2}}.
  \label{13}
\end{equation}
Substituting~\eqref{13} into~\eqref{12}, we can write $\tau_{5}$ as
\begin{equation}
  \tau_{5} = \left(\frac{\pi}{2}\right)^{1/3}v_{CY}^{1/3}
  \left(\frac{(\pi\rho)^{2}}
  {v_{CY}M_{Pl}}\right)^{2/3}.
  \label{14}
\end{equation}
For the reference parameters chosen in~\eqref{2} and~\eqref{3},
$\tau_{5}$ becomes
\begin{equation}
  \tau_{5} \sim 1.
  \label{15}
\end{equation}

The K\"ahler potential for the vector bundle moduli is less well-known.
It was shown in~\cite{Buchbinder:2003pi} that it always has the form
\begin{equation}
  \tilde{K}_{bundle}=kM_{Pl}^{2}K_{bundle}(\phi_{a}, \bar{\phi_{a}}),
  \label{16}
\end{equation}
where $k$ is a dimensionless constant given by
\begin{equation}
  k=\frac{1}{(4\pi)^{5/3}\Big((\pi\rho)^{2}M_{Pl}^{2}\Big)^{1/3}}
  \label{17}
\end{equation}
and $K_{bundle}$ is a dimensionless function of the vector bundle 
moduli $\phi_{a}$. For the reference parameters given in~\eqref{2} 
and~\eqref{3}, $k$ takes the value
\begin{equation}
  k \sim 10^{-5}.
  \label{18}
\end{equation}
The generic properties of $K_{bundle}$ relevant to moduli
stabilization were discussed in detail in~\cite{Buchbinder:2002wz}.

We conclude that
\begin{equation}
  K=K_{S,T}+K_{Z}+K_{5}+\tilde{K}_{bundle}
  \label{19}
\end{equation}
is the total K\"ahler potential for all the moduli.

\subsection*{Superpotentials}

There are three non-vanishing contributions to the superpotential for
the moduli. First consider the flux-induced superpotential. Let us
turn on a non-zero flux of the Neveu-Schwarz three-form $H$ on the
Calabi-Yau threefold. The presence of this non-zero flux generates a
superpotential for the $h^{2,1}$ moduli of the
form~\cite{Behrndt:2000zh, Becker:2002jj}
\begin{equation}
  W_{f}=M_{Pl}^{3}h_{1}\int_{X}{\tilde{H} \wedge \tilde{\Omega}},
  \label{20}
\end{equation}
where 
\begin{equation}
  h_{1}=\frac{1}{v_{CY}^{1/2}M_{Pl}^{3}}
  \label{21}
\end{equation}
and $\tilde{H}$ and $\tilde{\Omega}$ are both dimensionless. They have
been obtain from $H$ and $\Omega$ by scaling with respect to the
appropriate reference parameters. For the values chosen in~\eqref{2}
and~\eqref{3}, $h_{1}$ becomes
\begin{equation}
  h_{1} \sim 10^{-8}.
  \label{22}
\end{equation}
As discussed in~\cite{Buchbinder:2003pi}, the warping away from a
Calabi-Yau threefold due to the flux will be negligibly small if we
take
\begin{equation}
  \int_{C}{\tilde{H}} \ll 10^{5},
  \label{23}
\end{equation}
where $C$ is an appropriate three-cycle. Henceforth, we will always
choose the flux to satisfy this condition.

Second, let us turn on a gaugino condensate on the hidden
brane~\cite{Dine:1985rz, Kaplunovsky:1993rd, Brignole:1993dj,
  Lukas:1998hk, Nilles:1997cm, Horava:1996vs, Lalak:1997zu,
  Lukas:1997rb, Lukas:1999kt}.  As discussed
in~\cite{Buchbinder:2003pi}, this produces a superpotential for the
$S$, $T^{I}$ and $\bf Y \rm$ moduli given by
\begin{equation}
  W_{g}=M_{Pl}^{3}h_{2}\exp 
  \left[ -\epsilon\left(S - \alpha^{(2)}_{I}T^{I}
  + \tau_{5} \frac{{\bf Y \rm}^{2}}{\cal{T}}\right)\right],
  \label{24}
\end{equation}
where 
\begin{equation}
  h_{2} \sim \frac{1}{M_{Pl}v_{CY}^{1/2}(\pi\rho)}
  \left(\frac{\kappa_{11}}{4\pi}\right)^{2/3}
  \label{25}
\end{equation}
For the values of the reference parameters chosen in~\eqref{2}
and~\eqref{3},
\begin{equation}
  h_{2} \sim 10^{-6}.
  \label{26}
\end{equation}
The coefficient $\epsilon$ is related to the coefficient $b_{0}$ of
the beta-function and is given by
\begin{equation}
  \epsilon=\frac{6\pi}{b_{0}\alpha_{GUT}}.
  \label{27}
\end{equation}
For the $E_{8}$ gauge group of our hidden sector $b_{0}=90$. Taking
$\alpha_{GUT}$ to have its phenomenological value given in~\eqref{2},
we obtain
\begin{equation}
  \epsilon \sim 5.
  \label{28}
\end{equation}
The term $\alpha^{(2)}_{I}T^{I}$ is related to the tension of the
hidden brane~\cite{Lukas:1997fg}. Choosing the hidden sector vector
bundle to be trivial, we find that
\begin{equation}
  \alpha^{(2)}_{I}=-\frac{\pi\rho}{32\pi v_{CY}}
  \left(\frac{\kappa_{11}}{4\pi}\right)^{2/3}\int_{z_{R}}{\omega_{I}},
  \label{29}
\end{equation}
where $z_{R}$ is the curve Poincare dual to $\Tr R \wedge R$.  One can
estimate the order of magnitude of $\alpha^{(2)}_{I}$ by using the
reference parameters~\eqref{2} and~\eqref{3} in~\eqref{29}. The result
is that
\begin{equation}
  \alpha^{(2)}_{I} \sim \frac{1}{v_{CY}^{1/3}}\int_{z_{R}}{\omega_{I}}.
  \label{15b}
\end{equation}
The term $\tau_{5}$ was presented in~\eqref{11}. Recall,
using~\eqref{2} and~\eqref{3}, that
\begin{equation}
  \tau_{5} \sim 1.
  \label{32}
\end{equation}
The real part of 
\begin{equation}
  S-\alpha_{I}^{(2)}T^{I}+\tau_{5}\frac{{\bf Y \rm}^{2}}{{\cal{T}}}
  \label{33}
\end{equation}
represents the inverse square of the gauge coupling parameter on the
hidden brane, with the last two terms being threshold
corrections~\cite{Lukas:1998hk, Lukas:1997rb}.

The third contribution to the moduli superpotential arises from
worldsheet instantons, that is, strings wrapped on holomorphic curves
in the Calabi-Yau threefold. In our context, these are generated by
membranes stretching between branes. At long wavelength, such
configurations reduce to strings wrapping holomorphic curves. There
are three different types of membrane configurations that contribute
to the superpotential.
\begin{enumerate}
\item A membrane stretching between the two orbifold planes.
\item A membrane beginning on the observable sector plane and ending on 
  the five-brane.
\item A membrane beginning on the five-brane and ending on the 
  hidden sector plane.
\end{enumerate}
Let us begin with the first configuration, a membrane stretching
between the two orbifold planes and wrapped on an isolated holomorphic
curve ${\cal{C}}$. It was shown in~\cite{Witten:1999eg} that its
non-perturbative contribution to the superpotential has the structure
\begin{equation}
  W_{np}[{\cal{C}}] \sim 
  \Pfaff\big({\cal{D}}_{-}|_{\cal{C}}\big)
  \exp\big(-\tau \tilde{\omega}_{I}T^{I}\big),
  \label{34}
\end{equation}
where 
\begin{equation}
  \tau=\frac{T_{M}\pi\rho}{2}v_{CY}^{1/3}
  \label{34a}
\end{equation}
and
\begin{equation}
  \tilde{\omega}_{I}=
  \frac{1}{v_{CY}^{1/3}}\int_{{\cal{C}}}{\omega_{I}}.
  \label{34b}
\end{equation}
Note that 
\begin{equation}
  T_{M}\pi\rho=\frac{1}{2\pi\alpha'}.
  \label{34c}
\end{equation}
For the reference parameters in~\eqref{2} and~\eqref{3},
$\tau$ becomes
\begin{equation}
  \tau \sim 10^{2}. 
  \label{34d}
\end{equation}
Henceforth, $\tau$ will be assumed to be much greater than unity,
which is naturally the case.  The factor
\begin{equation}
  \Pfaff\big({\cal{D}}_{-}|_{{\cal{C}}}\big)
  \label{36}
\end{equation}
in~\eqref{34} is the Pfaffian of the chiral Dirac operator twisted by
the observable sector bundle pulled back to the curve ${\cal{C}}$,
see~\cite{Witten:1999eg}.  This factor has been explicitly calculated
in a number of contexts~\cite{Buchbinder:2002ic, Buchbinder:2002pr,
  Buchbinder:2002ji} and found to be a homogeneous polynomial of the
``transition'' moduli of the curve ${\cal{C}}$. Note that there can be
many isolated curves ${\cal{C}}$ in the Calabi-Yau threefold. It has
been demonstrated that, in some contexts, the sum of the
superpotential contributions from these curves vanishes
identically~\cite{Beasley:2003fx}.  However, in the context of this
paper this is not generically the case, as shown explicitly
in~\cite{Buchbinder:2002pr}.

Now consider the second configuration, that is, a membrane stretching
between the observable sector plane and the five-brane. In this case
the membrane wraps on the same curve as the five-brane, namely
$z_{5}$.  The contribution to the superpotential is very similar
to~\eqref{34}. One finds that
\begin{equation}
  W_{5}^{(1)} \sim \Pfaff\big({\cal{D}}_{-}|_{z_{5}}\big)e^{-\tau {\bf Y \rm}}
  \label{37}
\end{equation}
where $\tau$ is given in~\eqref{34a} and
$\Pfaff({\cal{D}})_{-}|_{z_{5}}$ is~\eqref{36} with ${\cal{C}}=z_{5}$.
Similarly, the third configuration, that is, a membrane stretching
from the five-brane to the hidden sector plane, contributes
\begin{equation}
  W_{5}^{(2)} \sim e^{-\tau({\cal{T}}-{\bf Y \rm})}
  \label{38}
\end{equation}
to the superpotential. Note that since we choose a trivial vector
bundle on the hidden sector orbifold, the Pfaffian is just unity.

\section{Moduli Stabilization}
\label{sec:modfix}

In a typical heterotic vacuum, the number of moduli is rather large.
For example, in the minimal heterotic standard
model~\cite{Braun:2005nv}, there is the dilaton $S$, $3=h^{1,1}$
moduli $T^{I}$, $3=h^{2,1}$ moduli $Z_{\alpha}$, $13$ vector bundle
moduli $\phi_{a}$ and the five-brane translation modulus ${\bf Y
  \rm}$. Other heterotic vacua often have far more moduli, especially
vector bundle moduli whose number can be of ${\cal{O}}(10^{2})$.
Therefore, to obtain an explicit analytic solution for the moduli
potential and its minima we must simplify the model while retaining
its essential properties.

\subsection*{$\mathbf{h^{1,1}=1}$ Case}

It was argued in~\cite{Buchbinder:2003pi} that one can consider only
one $h^{1,1}$ modulus and one vector bundle modulus, $T$ and $\phi$
respectively, without any loss of generality. One need not restrict
the number of $h^{2,1}$ moduli $Z_{\alpha}$. Therefore, we will, in
this subsection, assume that the spectrum of our vacuum consists of
the moduli $S$, $T$, $Z_{\alpha}$, $\phi$ and ${\bf Y \rm}$.

The K\"ahler potential of our simplified model is given by
\begin{equation}
  K=K_{S,T}+K_{Z}+K_{5}+{\tilde{K}}_{bundle},
  \label{39}
\end{equation}
where~\eqref{6} becomes
\begin{equation}
  K_{S,T}=-M_{Pl}^{2}\ln(S+\bar{S})-3M_{Pl}^{2}\ln(T+\bar{T}),
  \label{40}
\end{equation}
expression~\eqref{7} remains
\begin{equation}
  K_{Z}=-M_{Pl}^{2}\ln\left(-i\int_{X}{\Omega \wedge \bar{\Omega}}\right)
  \label{41}
\end{equation}
and~\eqref{10} simplifies to
\begin{equation}
  K_{5}=2M_{Pl}^{2}\tau_{5}
  \frac{({\bf Y \rm}+\bar{{\bf Y \rm}})^{2}}
  {(S+\bar{S})(T+\bar{T})}.
  \label{42}
\end{equation}
In~\eqref{40}, we have chosen the Calabi-Yau intersection number to be
\begin{equation}
  d_{111}=1
  \label{42aa}
\end{equation}
for simplicity. Any other choice of $d_{111}$ will give identical
equations and leave the conclusions unchanged. Note from~\eqref{3a}
and~\eqref{3b} that
\begin{equation}
  \RE T=R.
  \label{42a}
\end{equation}
Also, by definition
\begin{equation}
  0  \leq \RE{\bf Y \rm} \leq \RE T,
  \label{44}
\end{equation}
since the five-brane must be between the orbifold planes.  In the
following we will always assume that
\begin{equation}
  |\IM T| \ll 1,
  \label{43}
\end{equation}
which is required to ignore cross coupling between $T$ and $\phi$.
For $\tilde{K}_{bundle}$ we choose
\begin{equation}
  \tilde{K}_{bundle}=kM_{Pl}^{2}K_{bundle}, \quad k \sim 10^{-5},
  \label{45}
\end{equation}
where, when $\phi$ is less than unity,
\begin{equation}
  K_{bundle}=-p\ln\big(\phi+\bar{\phi}\big)
  \label{46}
\end{equation}
and $p$ is a dimensionless, positive constant. Expression~\eqref{46}
is the simplest function satisfying all the requirements specified
in~\cite{Buchbinder:2003pi}. However, one can choose $K_{bundle}$ to
be any other function satisfying these requirements without altering
the conclusions of this paper.

Now consider the superpotential in our simplified vacuum. It is given
by
\begin{equation}
  W=W_{f}+W_{g}+W_{np}+W_{5}^{(1)}+W_{5}^{(2)},
  \label{47}
\end{equation}
where~\eqref{20} remains
\begin{equation}
  W_{f}=M_{Pl}^{3}h_{1}\int_{X}{\tilde{H} \wedge \tilde{\Omega}}
  \,, \quad 
  h_{1} \sim 10^{-8}
  \label{48}
\end{equation}
and~\eqref{24} reduces to
\begin{equation}
  W_{g}=M_{Pl}^{3}h_{2}\exp\left[-\epsilon\left(S-\alpha^{(2)}T+
      \tau_{5}\frac{{\bf Y \rm}^{2}}{T}\right)\right]
  \label{49}
\end{equation}
with
\begin{equation}
  h_{2} \sim 10^{-6}, \quad \epsilon \sim 5, \quad \alpha^{(2)} \sim 1, 
  \quad \tau_{5} \sim 1.
  \label{50}
\end{equation}
Since $h^{1,1}=1$, the space $H_{2}(X)$ is spanned by a single curve
class $z$. Any other effective class is simply a positive integer
multiple of $z$. For simplicity, we will take ${\cal{C}} =z_{5} =z$ in
the following. Other choices for $\cal{C}$ and $z_{5}$ will not alter
the conclusions of this subsection. It follows that the
non-perturbative superpotential~\eqref{34} is now given by
\begin{equation}
  W_{np}=c_{1}M_{Pl}^{3}\phi^{d+1}e^{-\tau T},
  \label{51}
\end{equation}
where 
\begin{equation}
  \tau \sim 10^{2},
  \label{51a}
\end{equation}
we have restored its natural scale and $c_{1}$ is some dimensionless
coefficient of order unity. The Pfaffian, which must be a homogeneous
polynomial, is represented by the factor $\phi^{d+1}$. We will assume
that $d+1$ is sufficiently large. This is the case in explicit
examples~\cite{Buchbinder:2002pr}. Finally, expressions~\eqref{37}
and~\eqref{38} for the five-brane superpotentials become
\begin{equation}
  W_{5}^{(1)}=c_{2}M_{Pl}^{3}\phi^{d+1}e^{-\tau {\bf Y \rm}}
  \label{52}
\end{equation}
and
\begin{equation}
  W_{5}^{(2)}=c_{3}M_{Pl}^{3}e^{-\tau (T-{\bf Y \rm})}
  \label{53}
\end{equation}
respectively, where $c_{2}$ and $c_{3}$ are dimensionless coefficients
of order unity.

Having specified the complete K\"ahler and superpotentials, one can now
solve for the minimum of the moduli potential energy. Specifically, we
will show that the system of equations
\begin{equation}
  D_{F}W=0,
  \label{54}
\end{equation}
where $D W$ is the K\"ahler covariant derivative, has a solution in the
correct phenomenological range for each field
$F=S,T,Z_{\alpha},\phi,{\bf Y \rm}$.

To begin with, consider the equations
\begin{equation}
  D_{Z_{\alpha}}W=0.
  \label{55}
\end{equation}
Under the assumption that
\begin{equation}
  \big|W_{f}\big| ~\gg~ 
  \big|W_{g}\big|,\,
  \big|W_{np}\big|,\,
  \big|W_{5}^{(1)}\big|,\,
  \big|W_{5}^{(2)}\big|
  \label{56}
\end{equation}
it was argued in~\cite{Buchbinder:2003pi} that~\eqref{55} should have
non-trivial solutions that fix each modulus $Z_{\alpha}$. The argument
employed here is identical to that given in~\cite{Kachru:2003aw} to
obtain moduli stabilization in the Type IIB context.

The remaining equations, namely
\begin{equation}
  D_{S}W=0, \quad D_{T}W=0, \quad D_{{\bf Y \rm}}W=0, \quad D_{\phi}W=0
  \label{57}
\end{equation}
were solved in detail in~\cite{Buchbinder:2003pi}. Here we simply
state the results. Writing
\begin{equation}
  S=S_{1}+i S_{2}, \quad T=T_{1}+i T_{2}
  \, \quad  
  {\bf Y \rm}={\bf Y \rm}_{1}+i{\bf Y \rm}_{2}
  \,, \quad 
  \phi=r e^{i\theta}
  \label{58}
\end{equation}
as well as
\begin{equation}
  W_{f}=|W_{f}| \, e^{i f},
  \label{59}
\end{equation}
the solution to equations~\eqref{57} were found to be the following.
First,
\begin{equation} 
  S_{1} \sim 1, \quad S_{2} \sim -\frac{f+2\pi n_{1}}{\epsilon},
  \label{60}
\end{equation}
where $n_{1}$ is an arbitrary integer. Note that $\RE S \sim 1$, as
required by~\eqref{4}.  Second, one finds
\begin{equation} 
  T_{1} \sim 1, \quad T_{2} \sim -\frac{f+\pi(2n_{3}+1)}{\tau},
  \label{61}
\end{equation}
with $n_{3}$ an arbitrary integer. Again, $\RE T=R \sim 1$, as
required by~\eqref{4}.  Since naturally $\tau \gg 1$, one can choose
$n_{3}$ so that $|\IM T| \ll 1$ which is consistent with
assumption~\eqref{43}. Third,
\begin{equation} 
  {\bf Y \rm}_{1}= \frac{S_{1}}{2\tau_{5}}
  \,, \quad 
  {\bf Y \rm}_{2} \sim
  0
  \,.
  \label{62}
\end{equation}
Recalling that $\tau_{5} \sim 1$, we see that $\RE{\bf Y \rm} \le \RE
T$, consistent with~\eqref{44}. Finally, one finds that
\begin{equation}
  r=\left( \frac{pk|W_{f}|e^{\tau {\bf Y \rm}_{1}}}
    {2(d+1)c_{2}cos(\theta)}\right)^{1/d}, \quad  \theta=\frac{f+2\pi n_{2}}{d}.
  \label{64}
\end{equation}
Note that the five contributions to the superpotential, when evaluated
at these field values, satisfy the assumption~\eqref{56}, as they
must.

It is straightforward to find the value of the potential energy at
this minimum. It is given by
\begin{equation}
  V_{min}=-3e^{K/M_{Pl}^{2}}\frac{|W|^{2}}{M_{Pl}^{2}} \sim -\frac{|W_{f}|^{2}}{M_{Pl}^{2}}.
  \label{65}
\end{equation}
The size of the potential energy is, therefore, determined by the
flux-induced superpotential. Since from~\eqref{20} and~\eqref{22}
$W_{f}$ is of ${\cal{O}}(10^{-8}M_{Pl}^{3})$, we expect $V_{min}$ to
be
\begin{equation}
  V_{min} \sim -10^{-16}M_{Pl}^{4} \sim -10^{60}(GeV)^{4}.
  \label{66}
\end{equation}
Note that larger values of flux can be allowed, as long as they
satisfy the constraint~\eqref{23}. In this case the potential will
take an even larger negative value at the minimum. Finally, since
$D_{F}W=0$ in this vacuum, $N=1$ supersymmetry remains unbroken.

\subsection*{$\mathbf{h^{1,1}>1}$ Case}

As argued in~\cite{Buchbinder:2003pi}, there is no reason why the
above analysis cannot be extended to heterotic vacua with larger
numbers of K\"ahler and vector bundle moduli. In this section, we will
generalize the above discussion to include an arbitrary number of
K\"ahler moduli $T^{I}$, $I=1,\dots ,h^{1,1}$, one vector bundle modulus
$\phi$, any number of complex structure moduli $Z_{\alpha}$ and the
five-brane translation modulus ${\bf Y \rm}$. Vacua with more than one
vector bundle modulus will be considered elsewhere~\cite{toappear}.
For specificity, we will analyze our theory in the case of $h^{1,1}=2$
K\"ahler moduli, $T^{1}$ and $T^{2}$. It will be clear from the
discussion that this captures all relevant information and is easily
generalized to arbitrary $h^{1,1}$.

The K\"ahler potential of our vacuum is now 
\begin{equation}
  K=K_{S,T}+K_{Z}+K_{5}+{\tilde{K}}_{bundle},
  \label{67}
\end{equation}
where $K_{Z}$ and ${\tilde{K}}_{bundle}$ are given by~\eqref{7}
and~\eqref{46},\eqref{47} respectively.  To find $K_{S,T}$
from~\eqref{6} we must specify the intersection numbers $d_{IJK}$.
Since $h^{1,1}=2$, the spaces $H^{1,1}(X)$ and $H_{2}(X)$ are spanned
by $\{\omega_{1},\omega_{2}\}$ and its dual basis $\{z^{1},z^{2}\}$
respectively. $X$ being torus-fibered allows us to identify $z^{1}$
with the fiber class and $z^{2}$ with the curves in the base. Now,
note that taking the volume of $z^{1}$ to zero must make $V$, the
volume of $X$, vanish. It then follows from~\eqref{3b} that
\begin{equation}
  d_{222}=0.
  \label{67a}
\end{equation}
Similarly, letting the volume of $z^{2}$ vanish should not send $V$ to
zero. Hence, $d_{111} \neq 0$. Finally, any term involving $d_{122}$
would be sub-dominant in our analysis.  Hence, we can choose the
Calabi-Yau intersection numbers to vanish with the exception of
\begin{equation}
  d_{111}=1, \quad d_{112}=1
  \label{68}
\end{equation}
without loss of generality.  Intersection numbers of this type will
appear in the minimal heterotic standard model discussed below. It
follows that $K_{S,T}$ is given by~\eqref{6} with intersection
number~\eqref{68}.  To specify $K_{5}$, one must specify the curve
$z_{5}$ on which the five-brane is wrapped.  It is well-known [] that
wrapping a string over a fiber of a torus-fibration does not
contribute to the superpotential. For this reason, in this subsection
we will choose
\begin{equation}
  z_{5}=z^{1}+z^{2}.
  \label{68a}
\end{equation}
which has a component in the base.  It follows from~\eqref{5c} that
\begin{equation}
  {\cal{T}}=T^{1}+T^{2}.
  \label{68b}
\end{equation}
$K_{5}$ is then given by~\eqref{10} with ${\cal{T}}$ defined
in~\eqref{68b}. It is helpful to re-express $K_{S,T}$ in terms of
fields ${\cal{T}}$ and $T^{2}$. The second term in~\eqref{6} then
becomes
\begin{equation}
  K_{S,T}=\dots -M_{Pl}^{2}\ln\Big(({\cal{T}}+\bar{{\cal{T}}})^{3}
  -2({\cal{T}}+\bar{{\cal{T}}})^{2}
  (T^{2}+{\bar{T}}^{2})+({\cal{T}}+
  \bar{{\cal{T}}})(T^{2}+{\bar{T}}^{2})^{2}\Big).
  \label{68c}
\end{equation}

Similarly, the superpotential is now
\begin{equation}
  W=W_{f}+W_{g}+W_{np}+W_{5}^{(1)}+W_{5}^{(2)},
  \label{69}
\end{equation}
with $W_{f}$ and $W_{g}$ given in~\eqref{20} and~\eqref{24}
respectively. It is helpful to re-express the $\alpha_{I}^{(2)}T^{I}$
term in~\eqref{24} in terms of ${\cal{T}}$ and $T^{2}$.  It follows
that
\begin{equation}
  \alpha_{I}^{(2)}T^{I}=\alpha_{1}{\cal{T}}+c T^{2},
  \label{69A}
\end{equation}
where coefficient
\begin{equation}
  c=\epsilon(\alpha_{2}^{(2)}-\alpha_{1}^{(2)}).
  \label{69B}
\end{equation}
Hence, one can write $W_{g}$ as
\begin{equation}
  W_{g}=W_{g}|_{{\cal{T}}} \, e^{c T^{2}}.
  \label{69C}
\end{equation}
To present $W_{np}$ in~\eqref{34}, one must give the curves
${\cal{C}}$ contributing to this superpotential.  In general,
enumerating and specifying these curves is a very difficult problem
which has not been fully solved. In this subsection, we will simplify
the problem and take
\begin{equation}
  {\cal{C}}=z_{5}=z^{1}+z^{2}, \quad z^{2}.
  \label{69c}
\end{equation}
This choice reflects the fact that $h^{1,1}=2$ and that both $z_{5}$
and $z^{2}$ have a component in the base. Then~\eqref{34} becomes
\begin{equation}
  W_{np}=W_{np}|_{z_{5}}+W_{np}|_{z^{2}}.
  \label{69d}
\end{equation}
$W_{5}^{(1)}$ and $W_{5}^{(2)}$ are given by~\eqref{37} and~\eqref{38}
respectively.  As in the $h^{1,1}=1$ case, we continue to have
\begin{equation}
  k \sim 10^{-5},\quad  h_{1} \sim 10^{-8},\quad  h_{2} \sim 10^{-6},\quad \epsilon \sim 5,
  \quad \alpha^{(2)}_{1} \sim 1, \quad \tau_{5} \sim 1, \quad \tau \sim 10^{2}.
  \label{70}
\end{equation}

Having specified the K\"ahler and superpotentials, we now solve for the
minimum of the moduli potential energy. Specifically, we will show
that the system of equations
\begin{equation}
  D_{F}W=0,
  \label{71}
\end{equation}
where $D W$ is the K\"ahler covariant derivative, has a solution in the
correct phenomenological range for $F=S, {\cal{T}}, T^{2}, Z_{\alpha},
\phi, {\bf Y \rm}$.  We will denote
\begin{equation}
  {\cal{T}}={\cal{T}}_{1}+i{\cal{T}}_{2}, \quad T^{2}=t_{1}+it_{2}.
  \label{72}
\end{equation}
First consider the equations~\eqref{71} for $F=S,{\cal{T}},
Z_{\alpha}, \phi, {\bf Y \rm}$.  If we assume that
\begin{equation}
  t_{1} \lesssim {\cal{T}}_{1}, \quad |c|t_{1} \lesssim 1
  \,, \quad 
  \Big|W_{np}|_{z^{2}}\Big| \ll \big|W_{f}\big|,
  \label{74}
\end{equation}
then these equations are, to a good approximation, the same as in the
$h^{1,1}=1$ case.  Hence, the solutions for $S$, ${\cal{T}}$,
$Z_{\alpha}$, $\phi$, ${\bf Y \rm}$ remain essential those given
in~\eqref{60},~\eqref{61},~\eqref{55},~\eqref{64} and~\eqref{62}
respectively.  They become identical to them for $t_{1} \ll
{\cal{T}}_{1}$ and $|c|t_{1} \ll 1$.  Clearly, it greatly simplifies
our analysis if we can continue to use the $h^{1,1}=1$ results.  For
this reason, we will seek solutions under the assumption
that~\eqref{74} holds.

Now consider the $F=T^{2}$ equation
\begin{equation}
  \partial_{T^{2}} W=-\frac{1}{M_{Pl}^{2}}\big(\partial_{T^{2}}K\big) W.
  \label{75}
\end{equation}
Using~\eqref{64}, we find that the $W_{np}|_{z^{2}}$ term on the
left-hand side of~\eqref{75} can be neglected relative to $W_{g}$.
Then, assuming conditions~\eqref{74} are valid,~\eqref{75} becomes
\begin{equation}
  W_{g}|_{{\cal{T}}}e^{cT^{2}}c=\frac{1}{{\cal{T}}_{1}}W_{f}.
  \label{77}
\end{equation}
In the analysis of the $D_{S}W=0$ equation, one finds that
\begin{equation}
  W_{g}|_{{\cal{T}}}=-\frac{1}{2\epsilon S_{1}}W_{f}.
  \label{78}
\end{equation}
Substituting this into~\eqref{77} gives
\begin{equation}
  e^{cT^{2}}=-2\frac{\epsilon S_{1}}{c{\cal{T}}_{1}}.
  \label{79}
\end{equation}
Writing $T^{2}=t_{1}+it_{2}$, it follows that
\begin{equation}
  t_{1}=\frac{1}{c}\ln\left(\frac{2\epsilon S_{1}}{|c|{\cal{T}}_{1}}\right), 
  \label{80}
\end{equation}
where
\begin{equation}
  t_{2}=\frac{\pi(2n_{4}+1)}{c}, \quad t_{2}=\frac{2n_{4}\pi}{c}
  \label{80a}
\end{equation}
for $c>0$ and $c<0$ respectively and $n_{4}$ is any integer.
Let us parameterize
\begin{equation}
c= \pm\left(\frac{2\epsilon S_{1}}{{\cal{T}}_{1}}\right)^{1-\delta}, 
\label{80b}
\end{equation}
for $c>0$ and $c<0$ respectively.  Then, the solution for $t_{1}$
becomes
\begin{equation}
  t_{1}=|\delta|
  \left(\frac{2\epsilon S_{1}}{{\cal{T}}_{1}}\right)^{\delta-1}
  \ln\left(\frac{2\epsilon S_{1}}{{\cal{T}}_{1}} \right)
  \,,
  \label{80c}
\end{equation}
where $\delta > 0$ for $c>0$ and $\delta < 0$ when $c<0$. For
$|\delta| \lesssim 1$, $t_{1}$, $|c|t_{1}$ and $|W_{np}|_{z^{2}}|$
satisfy the assumptions in ~\eqref{74}.

As an example, recall from~\eqref{60},~\eqref{61} and~\eqref{70} that
$S_{1} \sim 1$, ${\cal{T}}_{1} \sim 1$ and $\epsilon \sim 5$. These
results were arrived at by implicitly choosing $|W_{f}|$ to be or order
$10^{-8}$. Then
\begin{equation}
  c=\pm 10^{1-\delta}
  \label{81}
\end{equation}
for $c>0$ and $c<0$ respectively and $t_{1}$
becomes
\begin{equation}
  t_{1}=|\delta|\big(10^{\delta-1}\ln(10)\big)
  \,,
  \label{82}
\end{equation}
where $\delta > 0$ for $c>0$ and $\delta < 0$ when $c<0$.  Note that
$\alpha_{2}^{(2)}-\alpha_{1}^{(2)} \sim .2$ when $|\delta|=1$ and
increases to $\alpha_{2}^{(2)}-\alpha_{1}^{(2)}=2$ for $|\delta| \ll
1$. These are realistic values for the type of Calabi-Yau threefolds
we are considering.  The general situation is the following. One
chooses a torus-fibered Calabi-Yau threefold with $h^{1,1}=2$ and
computes $\alpha_{2}^{(2)}-\alpha_{1}^{(2)}$. This is expected to be a
number of ${\cal{O}}(1)$. Then adjust the flux superpotential so that
\begin{equation}
1-\frac{|\alpha_{2}^{(2)}-\alpha_{1}^{(2)}|}{2S_{1}/{\cal{T}}_{1}} 
\sim {\cal{O}}(\delta).
\label{82a}
\end{equation}
In this way one fine-tunes the left-hand expression to give $\delta$ and, hence, $t_{1}$
of the desired value. We will explore this in more generality elsewhere.

We conclude that in the $h^{1,1}=2$ case, there exists a solution of
the equations $D_{F}W=0$ for $F=S,{\cal{T}},T^{2}, Z_{\alpha},
\phi,{\bf Y \rm}$. The values for $S$, ${\cal{T}}$, $Z_{\alpha}$,
$\phi$, ${\bf Y \rm}$ are those given in the $h^{1,1}=1$ case and
$T^{2}=t_{1}+it_{2}$, where
\begin{equation}
  t_{1}=|\delta|
  \left(\frac{2\epsilon S_{1}}{{\cal{T}}_{1}}\right)^{\delta-1}
  \ln\left(\frac{2\epsilon S_{1}}{{\cal{T}}_{1}} \right),
  \label{83}
\end{equation}
where $\delta > 0$ for $c>0$ and $\delta < 0$ when $c<0$ and
\begin{equation}
  t_{2}=\frac{\pi(2n_{4}+1)}{c}, \quad t_{2}=\frac{2n_{4}\pi}{c}
  \label{83b}
\end{equation}
for $c>0$ and $c<0$ respectively and $n_{4}$ is any integer.  Note
that since $t_{1} \lesssim {\cal{T}}_{1}$, it follows
from~\eqref{3a},~\eqref{3b} and~\eqref{68} that $\RE{\cal{T}} \sim R$.
Hence, we see from~\eqref{60} and~\eqref{61} that $V$ and $R$ continue
to be stabilized at the phenomenologically acceptable values of
\begin{equation}
  V \sim 1, \quad R \sim 1.
  \label{83a}
\end{equation}
Since $|W_{f}|$ continues to dominate the contributions to the
superpotential, we have
\begin{equation}
  V_{min} \sim -10^{-16}M_{Pl}^{4} \sim -10^{60}(GeV)^{4},
  \label{84}
\end{equation}
as in the $h^{1,1}$ case. Since $D_{F}W=0$ for all fields $F$, $N=1$
supersymmetry remains unbroken in this vacuum. 

We considered the case $h^{1,1}=2$ case for specificity only. The
analysis is easily repeated for any number of K\"ahler moduli $T^{I}$,
$I=1,\dots,h^{1,1}$ when $h^{1,1}>2$. The solutions are similar and
easily found and the same conclusions will hold.

\subsection*{Results}

The above results are based on our analysis of simplified, but
representative, heterotic $M$-theory vacua with a non-trivial,
slope-stable holomorphic vector bundle in the observable sector, a
trivial vector bundle without Wilson lines in the hidden sector and a holomorphic
five-brane in the bulk space. We found the following.
\begin{itemize}
\item For a natural range of parameters, the potential energy function
  of the moduli fields has a minimum which fixes the values of
  \emph{all} moduli. Both $\RE S$ and $R$ at this minimum are of
  ${\cal{O}}(1)$, as required on phenomenological grounds. The vacuum
  value of $\RE{\bf Y \rm}$ is between 0 and $R$, as it must be.
\item The potential energy function evaluated at this minimum is
  negative. Its value is typically large, of order $V_{min} \sim
  -10^{-16}M_{Pl}^{4}$. Hence, this theory has a \emph{large,
    negative} cosmological constant.
\item The K\"ahler covariant derivatives $D W$ vanish for all moduli
  fields at this minimum. Hence, supersymmetry is \emph{unbroken} in
  this vacuum, despite the occurrence of gaugino condensation in the
  hidden sector.
\end{itemize}
Despite the fact that these results were derived in a simplified
model, we see no reason that they do not apply to heterotic vacua with
larger numbers of vector bundle moduli, as argued
in~\cite{Buchbinder:2003pi}. We note that our conclusions are
consistent with the results of~\cite{Buchbinder:2004im}, in the
heterotic $M$-theory context, and~\cite{Dasgupta:1999ss,
  Giddings:2001yu, Kachru:2002he, Kachru:2002sk, Frey:2002hf,
  Gukov:1999ya, Taylor:1999ii, Curio:2000sc, Curio:2003ur,
  Balasubramanian:2004uy, Balasubramanian:2005zx}, in the context of
Type IIB superstrings. In the case of Type IIB strings, it was shown
in~\cite{Kachru:2003aw} that one could ``lift'' the minimum of the
moduli potential to a positive value by adding anti D-branes to the
theory. The new minimum continues to fix all moduli, has a
phenomenologically acceptable small positive cosmological constant,
albeit by fine-tuning~\cite{Weinberg:1987dv}, and, due to the presence
of anti D-branes, breaks supersymmetry.  As was discussed
in~\cite{Braun:2006ae}, a topologically stable configuration of
five-branes and anti-five-branes occurs naturally in the heterotic
standard models presented in~\cite{Braun:2005ux, Braun:2005bw,
  Braun:2005zv, Braun:2005nv}. It is reasonable, therefore, to ask
whether the existence of anti-five-branes in a heterotic $M$-theory
vacuum might, when combined with the results of this section, lead to
a vacuum with a small, positive cosmological constant that breaks
supersymmetry and fixes all moduli. The answer is affirmative.  This
is the content of the following section.

As a final comment, note that above we have solved the $D_{F}W=0$
equations under the assumption that the fields satisfy~\eqref{74}.
This allowed us to use the solutions for $S$, ${\cal{T}}$,
$Z_{\alpha}$, $\phi$ and ${\bf Y \rm}$ obtained in the $h^{1,1}=1$
case and, hence, to simplify the analysis. However, this assumption is
not necessary. If at least one of the conditions in~\eqref{74} is not
imposed, then one would have to search for new solutions for {\it all}
of the $D_{F}W=0$ equations. Since these can be continuously relaxed
to the equations discussed above, we see no reason why solutions would
not exist. That is, we expect there to be supersymmetric vacua with a
negative cosmological constant over a wide range of moduli space.
However, we have not verified this explicitly.

\section{Adding Anti-Five-Branes}
\label{sec:anti}

In~\cite{Kachru:2003aw}, it was argued that the vacuum state of
flux-compactifications with $D$-branes and gaugino condensation in
certain Type IIB theories completely fixes all moduli fields, albeit
with a large, negative cosmological constant and without breaking
supersymmetry.  It was then shown that introducing an anti D-brane
into such a vacuum adds a term proportional to the tension of the anti
$D$-brane to the original $N=1$ supersymmetric Lagrangian.  This
additional term was shown to lift the minimum to a meta-stable vacuum
that fixes all moduli, breaks the $N=1$ supersymmetry and has, after
fine-tuning, a phenomenologically acceptable small, positive
cosmological constant. This mechanism has recently been studied in
more detail in~\cite{Choi:2004sx, Choi:2005ge}, who reach the same
conclusion and explicitly compute the induced supersymmetry breaking
terms.

In the context of heterotic $M$-theory, we have seen in the previous
section that the vacuum state of flux-compactifications with
five-branes and gaugino condensation fixes all moduli, but with a
large, negative cosmological constant and without supersymmetry
breaking.  Following~\cite{Kachru:2003aw}, we now add an
anti-five-brane in the vacuum state. One can perform, in the heterotic
context, a calculation similar to that described
in~\cite{Kachru:2003aw}. This was carried out
in~\cite{Buchbinder:2004im} with the result that, again, the original
$N=1$ supersymmetric Lagrangian is modified by the addition of a term
proportional to the tension of the anti-five-brane. Specifically, this
term was found to be
\begin{equation}
  \Delta U_{\bar{5}}= \frac{4T_{5}}{V^{4/3}R^{2}} \int_{X}{\omega \wedge J},
  \label{85}
\end{equation}
where
\begin{equation}
  J=c_{2}(V)-c_{2}(TX)+[{\cal{W}}]+[\bar{{\cal{W}}}],
  \label{87}
\end{equation}
the integral is with respect to the K\"ahler form
\begin{equation}
  \omega =a^{I}\omega_{I}
  \label{88}
\end{equation}
and
\begin{equation}
  T_{5}=(2\pi)^{1/3}\left(\frac{1}{2\kappa_{11}^{2}}\right)^{2/3}, \quad \kappa_{11}^{2}=
  \frac{\pi\rho v_{CY}}{M_{Pl}^{2}}.
  \label{89}
\end{equation}
In~\eqref{87}, $[{\cal{W}}]$ is the cohomology class of the wrapped
five-brane, $[\bar{{\cal{W}}}]$ is the class of the wrapped
anti-five-brane and we have used the fact that the hidden sector
bundle $V'$ has been chosen to be trivial. The first two terms
in~\eqref{87} arise from the ``end-of-the-world'' orbifold planes. The
third and fourth terms are generated by the five-brane and
anti-five-brane respectively. Quantum consistency demands that the
theory be anomaly free. It follows that one must require that
\begin{equation}
  c_{2}(V)-c_{2}(TX)+[{\cal{W}}]-[\bar{{\cal{W}}}]=0
  \label{90}
\end{equation}
and, hence,~\eqref{87} becomes
\begin{equation}
  J=2[\bar{{\cal{W}}}].
  \label{91}
\end{equation}
It is important to note that if there was no anti-five-brane in the
vacuum J and, hence, $\Delta U_{\bar{5}}$ would vanish. However, in
the presence of an anti-five-brane one has a non-zero addition to the
potential energy given by
\begin{equation}
  \Delta U_{\bar{5}}= 8T_{5}{\cal{V}}_{\bar{5}}\frac{1}{V^{4/3}R^{2}},
  \label{92}
\end{equation}
where
\begin{equation}
  {\cal{V}}_{\bar{5}}= \int_{z_{\bar{5}}}{\omega}
  \label{93}
\end{equation}
is the volume of the curve $z_{\bar{5}}$ on which the anti-five-brane
is wrapped.

Given~\eqref{92}, one can add it to the $N=1$ supersymmetric
Lagrangian discussed in the previous section, find the new equations
of motion and re-solve for the vacuum solution.  We find results
completely consistent with those presented in~\cite{Buchbinder:2004im}
and~\cite{Kachru:2003aw} for the Type IIB string. First, the theory
continues to possess a local minimum that fixes all moduli.  It is
related to the supersymmetric vacuum found in the previous section in
the sense that it continuously relaxes to it as $\Delta U_{\bar{5}}$
is switched off. Second, the values of the moduli at this minimum are
not significantly altered over those in the supersymmetric vacuum. In
particular, one still finds that
\begin{equation}
  V \sim 1, \quad R \sim 1.
  \label{93a}
\end{equation}
Third, the addition of $\Delta U_{\bar{5}}$ to the Lagrangian provides
a source of positive energy, uplifting the value of the potential at
the minimum. Comparing~\eqref{92} and~\eqref{93} with~\eqref{66}, and
using~\eqref{93a}, we see that the cosmological constant at the
uplifted vacuum can be made positive with an arbitrarily small value
as long as one chooses
\begin{equation}
  T_{5}{\cal{V}}_{\bar{5}} \sim 10^{-16}GeV
  \label{94}
\end{equation}
or, equivalently, using~\eqref{2},~\eqref{3} and~\eqref{89} that
\begin{equation}
  \frac{{\cal{V}}_{\bar{5}}}{v_{CY}^{1/3}} \sim 10^{-7}.
  \label{95}
\end{equation}
Four, $N=1$ supersymmetry is broken in the uplifted vacuum. Fifth, as
in the Type IIB theories studied in~\cite{Kachru:2003aw}, this
uplifted vacuum is now meta-stable, with a possibility to tunnel to
large values of $\RE S$ and $\RE {\cal T}$. The lifetime of this
meta-stable vacuum, however, will be very long. And sixth, recall that
$Re{\cal{T}} \sim 1$ for any $h^{1,1} \geq 1$. It then follows
from~\eqref{3a},~\eqref{5d}, \eqref{88}, and~\eqref{93a} that
\begin{equation}
  \frac{{\cal{V}}_{{5}}}{v_{CY}^{1/3}} \sim 1.
  \label{95a}
\end{equation}
All of the above properties are straightforward, if tedious, to prove
and are very similar to the calculations required in the Type IIB
case.  Hence, we won't discuss them further here. A detailed analysis
of the supersymmetry breaking terms along the lines
of~\cite{Choi:2004sx, Choi:2005ge} will be presented elsewhere. What
is necessary for us to prove here is that the conditions~\eqref{95}
and~\eqref{95a} can consistently be achieved in our context.

Let us show that this is possible for the $h^{1,1}=2$ case discussed
in the previous section.  Recall that $z_{5}=z^{1}+z^{2}$ is the curve
on which the five-brane is wrapped. Then, its volume is given by
\begin{equation}
  {\cal{V}}_{5}=\int_{z_{5}=z^{1}+z^{2}}{\omega}.
  \label{96}
\end{equation}
Using~\eqref{5b} and~\eqref{88} we see that
\begin{equation}
  \frac{{\cal{V}}_{5}}{v_{CY}^{1/3}}=a^{1}+a^{2}.
  \label{96a}
\end{equation}
Now assume that there is an anti five-brane added to the theory which
is wrapped on the curve $z^{2}$. Then its volume is
\begin{equation}
  {\cal{V}}_{\bar{5}}=\int_{z^{2}}{\omega}
  \label{96b}
\end{equation}
which, using~\eqref{5b} and~\eqref{88}, becomes
\begin{equation}
  \frac{{\cal{V}}_{\bar{5}}}{v_{CY}^{1/3}}=a^{2}.
  \label{96c}
\end{equation}
In this vacuum, $t_{1}$ and, hence, $a^{2}$ are much
smaller than unity. It then follows from~\eqref{3b} that $a^{1}\sim
V^{1/3}$. Recalling that $V \sim 1$, expression~\eqref{96a} becomes
\begin{equation}
  \frac{{\cal{V}}_{5}}{v_{CY}^{1/3}} \sim 1.
  \label{97}
\end{equation}
Now note from~\eqref{83} and~\eqref{93a} that $a^{2} \sim
{\cal{O}}(\delta)$. By fine-tuning $\delta$ to be small, one can set
\begin{equation}
  \frac{{\cal{V}}_{\bar{5}}}{v_{CY}^{1/3}} \sim 10^{-7}
  \,,
  \label{97small}
\end{equation}
as desired. Thus, in the $h^{1,1}=2$ vacuum discussed above,
conditions~\eqref{95} and~\eqref{95a} can be simultaneously satisfied.

In a general vacuum, to obtain a small,
positive cosmological constant, one must work in a region of K\"ahler
moduli space in which~\eqref{95} and~\eqref{95a} are simultaneously
satisfied. Furthermore, assuming that such a region exists, it must be
demonstrated that the observable sector vector bundle $V$ is
slope-stable with respect to at least one such K\"ahler modulus.
Fortunately, all of these conditions can be simultaneously satisfied.
To demonstrate this requires that one give an explicit vacuum. In
particular, it is necessary to present the five-brane curve $z_{5}$,
the anti-five-brane curve $z_{\bar{5}}$, a holomorphic vector bundle
$V$ on the observable sector and the explicit region of the K\"ahler
cone for which $V$ is slope-stable. We will do this in the next
section for the minimal heterotic standard model.

\section{Minimal Heterotic Standard Model}
\label{sec:dS}

The minimal heterotic standard model was presented
in~\cite{Braun:2005nv}. The observable sector consists of a
holomorphic vector bundle $V$ with structure group $SU(4)$ over a
torus-fibered Calabi-Yau threefold $X$ with fundamental group
${\mathbb Z}_{3} \times{\mathbb Z}_{3}$.  This leads to a low energy
theory whose matter content in the observable sector is exactly that
of the MSSM. As discussed in~\cite{Braun:2006ae}, it is easiest to
choose a trivial bundle $V'={\cal{O}}_{X}$ in the hidden sector. The
number and properties of five-branes and anti-five-branes in the bulk
space are then determined by the requirement that the theory be
anomaly free.

\subsection*{Minimal Heterotic Standard Model}

The Calabi-Yau threefold $X$ is constructed from a covering space
$\tilde{X}$ on which ${\mathbb Z}_{3} \times{\mathbb Z}_{3}$ acts
freely. Then
\begin{equation}
  X=\tilde{X} \Big/ \big( \mathbb{Z}_3 \times\mathbb{Z}_3 \big).
  \label{98}
\end{equation}
Denote the quotient map as $q:{\tilde{X}} \rightarrow X$.  It was
shown in~\cite{Braun:2004xv} that $H^{1,1}({\tilde{X}})^{{\mathbb
    Z}_{3} \times{\mathbb Z}_{3}}$ is three-dimensional and spanned by
the cohomology classes $\tau_{1}$, $\tau_{2}$ and $\phi$.  It follows
that $H^{1,1}(X)$ is also three-dimensional and spanned by
$\omega_{I}$, $I=1,2,3$ where $q^{*}\omega_{1}=\tau_{1}$,
$q^{*}\omega_{2}=\tau_{2}$ and $q^{*}\omega_{3}=\phi$.  The
intersection numbers of $X$ are defined to be
\begin{equation}
  d_{IJK}=\frac{1}{v_{CY}}\int_{X}{\omega_{I} \wedge \omega_{J} \wedge \omega_{K}.}
  \label{98a}
\end{equation}
By pulling this expression back to $\tilde{X}$, one can compute these
coefficients. We find that
\begin{equation}
  \begin{gathered}
  d_{112}=d_{121}=d_{211}=\frac{1}{3}
  \,, \quad 
  d_{122}=d_{212}=d_{221}=\frac{1}{3}
  \,, \\
  d_{123}=d_{132}=d_{213}=d_{231}=d_{312}=d_{321}=1
  \,.    
  \end{gathered}
  \label{98b}
\end{equation}
Note that these intersection numbers are mathematically similar to the
those presented in~\eqref{68}. Specifically, they imply that
$z^{1}$,$z^{2}$ and $z^{3}$ in the minimal heterotic standard model
correspond to $z^{1}$ and $z^{2}$ in the $h^{1,1}=2$ case,
respectively. There are certain codimension one boundaries in
$H^{1,1}(X)$ where geometric transitions occur, that is, where the
volume of at least one curve vanishes. To exclude these regions, one
must choose a K\"ahler form in the three-dimension K\"ahler
cone~\cite{Gomez:2005ii}
\begin{equation}
  {\cal{K}}=\Big\{ a^{1}\omega_{1}+a^{2}\omega_{2}+a^{3}\omega_{3} 
  \Big| a^{1},a^{2},a^{3}> 0 \Big\} 
  \subset H^{1,1}(X).
  \label{99}
\end{equation}
Henceforth, we will consider only K\"ahler classes
\begin{equation}
  \omega=a^{1}\omega_{1}+a^{2}\omega_{2}+a^{3}\omega_{3} \in {\cal{K}}.
  \label{100}
\end{equation}
Recall from~\eqref{3b} that $V=\frac{1}{6}d_{IJK}a^{I}a^{J}a^{K}$. It
follows from~\eqref{99} that
\begin{equation}
  V=\frac{1}{6}\Big(
  \big(a^{1}\big)^{2}a^{2}+a^{1}\big(a^{2}\big)^{2}+
  6a^{1}a^{2}a^{3}\Big).
  \label{101}
\end{equation}
\begin{figure}[htbp]
  \centering
  \input{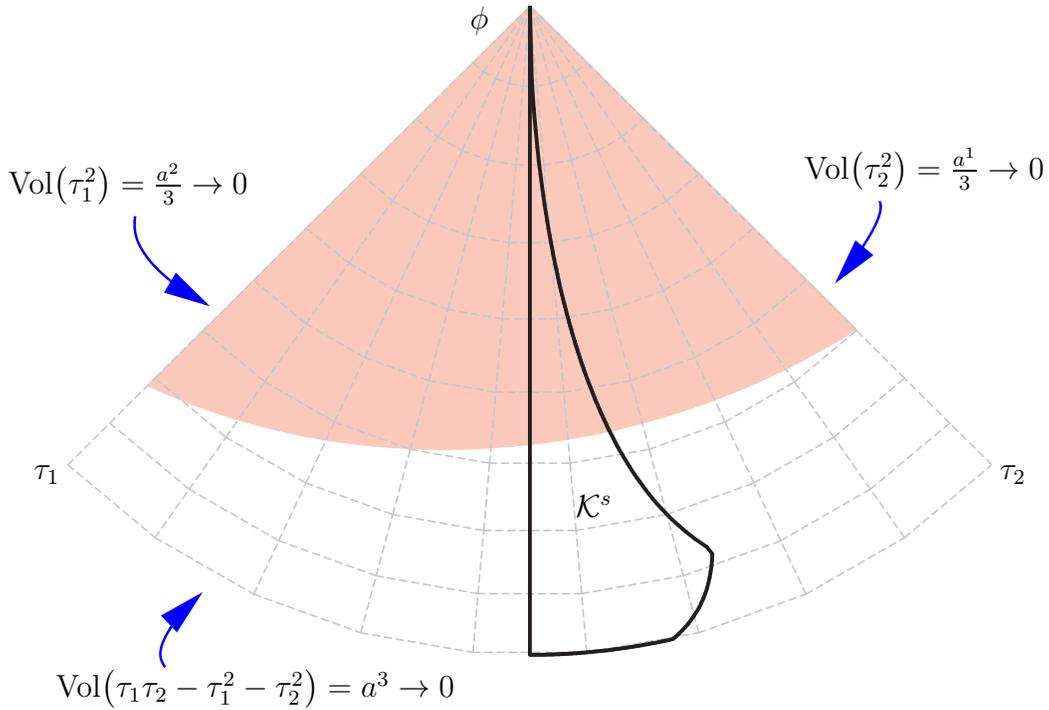}
  \caption{K\"ahler Cone. The observable sector vector bundle is
    slope-stable in the region $\Kcone^s$.}
  \label{fig:kahlercone}
\end{figure}
The slope-stability of the observable sector vector bundle $V$ was
proven in~\cite{Braun:2006ae}.  Specifically, it was shown that $V$ is
slope-stable with respect to any K\"ahler modulus in a restricted
three-dimensional region ${\cal{K}}^s$ of the K\"ahler cone ${\cal{K}}$.
As will be important in the following, one can enlarge the allowed
region ${\cal{K}}^s$, see Appendix~\ref{sec:stability}. This enlarged
region is shown in Figure~\ref{fig:kahlercone}.

\subsection*{Volumes of Curves}

It was shown in~\cite{Braun:2005nv} that
\begin{equation}
  c_{2}(V)=\omega_{1}^{2}+4\omega_{2}^{2}+4\omega_{1}\omega_{2},                                 
  \quad c_{2}(TX)=12(\omega_{1}^{2}+\omega_{2}^{2}).
  \label{102}
\end{equation}
The cancellation of quantum anomalies requires that
\begin{equation}
  c_{2}(V)-c_{2}(TX)+[{\cal{W}}]-[\bar{{\cal{W}}}]=0.
  \label{103}
\end{equation}
where we have used the fact that $V'={\cal{O}}_{X}$.
Inserting~\eqref{99}, this condition becomes
\begin{equation}
  [{\cal{W}}]-[\bar{{\cal{W}}}]=
  11\omega_{1}^{2}+8\omega_{2}^{2}-4\omega_{1}\omega_{2}
  =\big(3\omega_{1}^{2}\big)+4\big(\omega_{1}^{2}+\omega_{2}^{2}\big)
  -4\big(\omega_{1}\omega_{2}-\omega_{1}^{2}-\omega_{2}^{2}\big)
  \,.
  \label{104}
\end{equation}
Note that the terms in brackets are Poincare dual to effective curves
on $X$. Since they appear with positive and negative coefficients the
overall curve is not effective, and we require a non-vanishing
anti-five-brane. The simplest way to cancel the anomaly is to set
\begin{equation}
  [{\cal{W}}]=7\omega_{1}^{2}+4\omega_{2}^{2}
  \, \quad 
  [\bar{{\cal{W}}}]= 4\Big(\omega_{1}
  \omega_{2}-\omega_{1}^{2}-\omega_{2}^{2}\Big), 
  \label{105}
\end{equation}
which we will do henceforth.

The volumes of the five-brane curve $z_{5}$ and the anti-five-brane
curve $z_{\bar{5}}$ are easily computed. First consider the
anti-five-brane curve. Note that
\begin{equation}
  \frac{{\cal{V}}_{\bar{5}}}{v_{CY}^{1/3}}=\frac{1}{v_{CY}^{1/3}}\int_{z_{\bar{5}}}{\omega}
  =\frac{1}{v_{CY}}\int_{X}{\omega \wedge [{\bar{\cal{W}}}]}.
  \label{106}
\end{equation}
It then follows from~\eqref{98b},~\eqref{100} and~\eqref{105} that 
\begin{equation}
  \frac{{\cal{V}}_{\bar{5}}}{v_{CY}^{1/3}}=4a^{3}.
  \label{107}
\end{equation}
Similarly, for the five-brane curve
\begin{equation}
  \frac{{\cal{V}}_{5}}{v_{CY}^{1/3}}=\frac{1}{v_{CY}^{1/3}}\int_{z_{5}}{\omega}
  =\frac{1}{v_{CY}}\int_{X}{\omega \wedge [{\cal{W}}]}
  \label{108}
\end{equation}
which, using~\eqref{98b},~\eqref{100} and~\eqref{105}, becomes
\begin{equation}
  \frac{{\cal{V}}_{5}}{v_{CY}^{1/3}}=\frac{4}{3}a^{1}+\frac{7}{3}a^{2}.
  \label{109}
\end{equation}
Now consider the region ${\cal{K}}^s$ in Figure 1. Note that as one
approaches the bottom of ${\cal{K}}^s$ the modulus $a^{3} \rightarrow
0$.  It follows from~\eqref{107} that
$\frac{{\cal{V}}_{5}}{v_{CY}^{1/3}}$ can be made arbitrarily small. In
particular, for the appropriate value of $a^{3}$ one
can set
\begin{equation}
  \frac{{\cal{V}}_{\bar{5}}}{v_{CY}^{1/3}} \sim 10^{-7},
  \label{110}
\end{equation}
as required by~\eqref{95}. Now note that the region ${\cal{K}}^s$ is
bounded on the left by the vertical line defined by $a^{1}=a^{2}$. For
very small values of $a^{3}$, it follows from~\eqref{101} and the fact
that $V \sim 1$ in this vacuum that $a^{1} \cong a^{2} \sim
(3)^{1/3}$. Therefore, for moduli inside of ${\cal{K}}^s$ near the
vertical line as $a^{3} \rightarrow 0$, one finds
\begin{equation}
  \frac{{\cal{V}}_{5}}{v_{CY}^{1/3}} \sim 1,
  \label{111}
\end{equation}
as required by~\eqref{95a}.

\subsection*{Results}

For the minimal heterotic standard model we have shown the following.
\begin{itemize}
\item Taking the hidden sector vector bundle to be trivial, the
  anomaly cancellation condition specifies that this vacuum has both a
  five-brane and an anti-five-brane in the $S^{1}/{\mathbb Z}_{2}$
  interval and uniquely fixes their cohomology classes.
\item Neglecting the anti-five-brane, this vacuum has the structure of
  the $h^{1,1}>1$ theories analyzed in Section 3. As a consequence,
  neglecting the anti-five-brane, all moduli are stabilized, but at an
  $N=1$ supersymmetry preserving minimum with $V_{min} \sim
  -10^{-16}M_{Pl}^{4}$.
\item Adding the anti-five-brane lifts the minimum to a meta-stable
  vacuum with a \emph{positive} cosmological constant.  The moduli are
  fixed in this vacuum and have phenomenologically acceptable values.
\item There is a region of the K\"ahler cone for which the
  cosmological constant has its experimental value and for which the
  observable sector vector bundle is \emph{slope-stable}. One expects
  that the K\"ahler moduli of the meta-stable vacuum can be fine-tuned
  to lie in this region.
\end{itemize}

%%%%%%%%%%%%%%%%%%%%%%%%%%%%%%%%%%%%%%%%%% 
\section*{Acknowledgments}

We are grateful to E.~Buchbinder, R.~Donagi, B.~Nelson, T.~Pantev, and M.~Schulz for
enlightening discussions. This research was supported in part by the
Department of Physics and the Math/Physics Research Group at the
University of Pennsylvania under cooperative research agreement
DE-FG02-95ER40893 with the U.~S.~Department of Energy and an NSF
Focused Research Grant DMS0139799 for ``The Geometry of
Superstrings.''
%%%%%%%%%%%%%%%%%%%%%%%%%%%%%%%%%%%%%%%%%% 

\appendix

\section{Improved Stability Bound}
\label{sec:stability}

Showing that the vector bundle $V$ on $X$ is stable is equivalent to
showing that $\widetilde{V}$ on $\Xt$ is equivariantly stable. This
can be done by finding a set of inequalities that the K\"ahler moduli
have to satisfy~\cite{Gomez:2005ii}, and then showing that a common
solution exist. For the minimal heterotic standard model, this has
been worked out in~\cite{Braun:2006ae}.

To find the potentially destabilizing sub-bundles one has to
decide if there are maps
\begin{subequations}
  \begin{gather}
    \label{eq:HomA}
    \Hom\Big( \oB1(-4t+2f),~ \oB1(-t+f)\otimes I_3 \Big)
    = 
    \Hom\Big( \oB1,~ \oB1(3t-f)\otimes I_3 \Big)
    \,,
    \\
    \label{eq:HomB}
    \Hom\Big( \oB2(-4t+2f),~ \oB2(-t+f)\otimes I_6 \Big)
    = 
    \Hom\Big( \oB2,~ \oB2(3t-f)\otimes I_6 \Big)
    \,.
  \end{gather}
\end{subequations}
If there are maps in eq.~\eqref{eq:HomA}, then the line bundles
$\oXt(-4\tau_1+\tau_2+2f)$ can map to $\Vt$, and hence must have
negative slope (or $\Vt$ is rendered unstable). On the other hand, if
there is no map then $\oXt(-4\tau_1+\tau_2+2f)$ imposes no restriction
on the stability of $\Vt$. In the proof of slope-stability presented
in~\cite{Braun:2006ae}, it was always assumed that there are maps of
the form eqns.~\eqref{eq:HomA},~\eqref{eq:HomB}. This leads to
correct, but too strong, inequalities for the K\"ahler moduli.

In fact, there are no maps in
eqns.~\eqref{eq:HomA},~\eqref{eq:HomB}. Recall that there is a single
section of $\oB{i}(3t-f)$, 
\begin{equation}
  H^0\Big( \Xt,\, \oB{i}(3t-f)\Big) = \C
  \,,
\end{equation}
unique up to scale. Its zero locus are $9$ disjoint sections
$\IP1\subset B_i$, permuted by the $\ZZZ$ action. Furthermore, the
points in the ideal sheaves $I_3$, $I_6$ are the singular points of
$I_1$ Kodaira fibers, which do not meet any smooth
sections. Therefore,
\begin{equation}
  \Hom\Big( \oB1,~ \oB1(3t-f)\otimes I_3 \Big)
  =
  \Hom\Big( \oB2,~ \oB2(3t-f)\otimes I_6 \Big)
  =
  0
  \,.
\end{equation}
Using this result, the Proposition~1 in~\cite{Braun:2006ae} can be
strengthened to
\begin{proposition}
  \label{prop:stable}
  If all line bundles $\oXt(a_1\tau_1+a_2\tau_2+b\phi)$ with
  \begin{multline}
    (a_1,a_2,b) \in \Big\{
    ( 1,-1,-1),\,
    (-1, 1,-1),\,
    (-2, 2, 0),\,
    ( 2,-2,-1),\,
    \\
    ( 2,-5, 1),\,
    ( 1,-4, 1),\,
    (-4, 1, 1)
    \Big\}
  \end{multline}
  have negative slope then the vector bundle $\Vt$
  is equivariantly stable.
\end{proposition}
The set $\Kcone^s$ of K\"ahler moduli such that all slopes are indeed
negative is shown in Figure~\ref{fig:kahlercone}.

\bibliographystyle{JHEP} \renewcommand{\refname}{Bibliography}
\addcontentsline{toc}{section}{Bibliography} \bibliography{uplift}

\providecommand{\href}[2]{#2}\begingroup\raggedright\begin{thebibliography}{10}

\bibitem{Braun:2005ux}
V.~Braun, Y.-H. He, B.~A. Ovrut, and T.~Pantev, {\it A heterotic standard
  model},  {\em Phys. Lett.} {\bf B618} (2005) 252--258,
  [\href{http://xxx.lanl.gov/abs/hep-th/0501070}{{\tt hep-th/0501070}}].

\bibitem{Braun:2005bw}
V.~Braun, Y.-H. He, B.~A. Ovrut, and T.~Pantev, {\it A standard model from the
  e(8) x e(8) heterotic superstring},  {\em JHEP} {\bf 06} (2005) 039,
  [\href{http://xxx.lanl.gov/abs/hep-th/0502155}{{\tt hep-th/0502155}}].

\bibitem{Braun:2005zv}
V.~Braun, Y.-H. He, B.~A. Ovrut, and T.~Pantev, {\it Vector bundle extensions,
  sheaf cohomology, and the heterotic standard model},
  \href{http://xxx.lanl.gov/abs/hep-th/0505041}{{\tt hep-th/0505041}}.

\bibitem{Braun:2005nv}
V.~Braun, Y.-H. He, B.~A. Ovrut, and T.~Pantev, {\it The exact mssm spectrum
  from string theory},  \href{http://xxx.lanl.gov/abs/hep-th/0512177}{{\tt
  hep-th/0512177}}.

\bibitem{Ovrut:2006yr}
B.~A. Ovrut, {\it A heterotic standard model},  {\em AIP Conf. Proc.} {\bf 805}
  (2006) 236--239.

\bibitem{Buchmuller:2005jr}
W.~Buchmuller, K.~Hamaguchi, O.~Lebedev, and M.~Ratz, {\it The supersymmetric
  standard model from the heterotic string},
  \href{http://xxx.lanl.gov/abs/hep-ph/0511035}{{\tt hep-ph/0511035}}.

\bibitem{Bouchard:2005ag}
V.~Bouchard and R.~Donagi, {\it An su(5) heterotic standard model},  {\em Phys.
  Lett.} {\bf B633} (2006) 783--791,
  [\href{http://xxx.lanl.gov/abs/hep-th/0512149}{{\tt hep-th/0512149}}].

\bibitem{Bouchard:2006dn}
V.~Bouchard, M.~Cvetic, and R.~Donagi, {\it Tri-linear couplings in an
  heterotic minimal supersymmetric standard model},
  \href{http://xxx.lanl.gov/abs/hep-th/0602096}{{\tt hep-th/0602096}}.

\bibitem{Greene:1986bm}
B.~R. Greene, K.~H. Kirklin, P.~J. Miron, and G.~G. Ross, {\it A three
  generation superstring model. 1. compactification and discrete symmetries},
  {\em Nucl. Phys.} {\bf B278} (1986) 667.

\bibitem{Greene:1986jb}
B.~R. Greene, K.~H. Kirklin, P.~J. Miron, and G.~G. Ross, {\it A three
  generation superstring model. 2. symmetry breaking and the low-energy
  theory},  {\em Nucl. Phys.} {\bf B292} (1987) 606.

\bibitem{Greene:1986ar}
B.~R. Greene, K.~H. Kirklin, P.~J. Miron, and G.~G. Ross, {\it A superstring
  inspired standard model},  {\em Phys. Lett.} {\bf B180} (1986) 69.

\bibitem{Blumenhagen:2005zh}
R.~Blumenhagen, G.~Honecker, and T.~Weigand, {\it Non-abelian brane worlds: The
  open string story},  \href{http://xxx.lanl.gov/abs/hep-th/0510050}{{\tt
  hep-th/0510050}}.

\bibitem{Blumenhagen:2006ux}
R.~Blumenhagen, S.~Moster, and T.~Weigand, {\it Heterotic gut and standard
  model vacua from simply connected calabi-yau manifolds},
  \href{http://xxx.lanl.gov/abs/hep-th/0603015}{{\tt hep-th/0603015}}.

\bibitem{Andreas:2004ja}
B.~Andreas and D.~Hernandez~Ruiperez, {\it U(n) vector bundles on calabi-yau
  threefolds for string theory compactifications},
  \href{http://xxx.lanl.gov/abs/hep-th/0410170}{{\tt hep-th/0410170}}.

\bibitem{Braun:2004xv}
V.~Braun, B.~A. Ovrut, T.~Pantev, and R.~Reinbacher, {\it Elliptic calabi-yau
  threefolds with z(3) x z(3) wilson lines},  {\em JHEP} {\bf 12} (2004) 062,
  [\href{http://xxx.lanl.gov/abs/hep-th/0410055}{{\tt hep-th/0410055}}].

\bibitem{Gomez:2005ii}
T.~L. Gomez, S.~Lukic, and I.~Sols, {\it Constraining the kaehler moduli in the
  heterotic standard model},
  \href{http://xxx.lanl.gov/abs/hep-th/0512205}{{\tt hep-th/0512205}}.

\bibitem{Braun:2006ae}
V.~Braun, Y.-H. He, and B.~A. Ovrut, {\it Stability of the minimal heterotic
  standard model bundle},  \href{http://xxx.lanl.gov/abs/hep-th/0602073}{{\tt
  hep-th/0602073}}.

\bibitem{Langacker:2004xy}
P.~Langacker, {\it Neutrino physics (theory)},  {\em Int. J. Mod. Phys.} {\bf
  A20} (2005) 5254--5265, [\href{http://xxx.lanl.gov/abs/hep-ph/0411116}{{\tt
  hep-ph/0411116}}].

\bibitem{Giedt:2005vx}
J.~Giedt, G.~L. Kane, P.~Langacker, and B.~D. Nelson, {\it Massive neutrinos
  and (heterotic) string theory},  {\em Phys. Rev.} {\bf D71} (2005) 115013,
  [\href{http://xxx.lanl.gov/abs/hep-th/0502032}{{\tt hep-th/0502032}}].

\bibitem{Braun:2005fk}
V.~Braun, Y.-H. He, B.~A. Ovrut, and T.~Pantev, {\it Heterotic standard model
  moduli},  {\em JHEP} {\bf 01} (2006) 025,
  [\href{http://xxx.lanl.gov/abs/hep-th/0509051}{{\tt hep-th/0509051}}].

\bibitem{Witten:2001bf}
E.~Witten, {\it Deconstruction, g(2) holonomy, and doublet-triplet splitting},
  \href{http://xxx.lanl.gov/abs/hep-ph/0201018}{{\tt hep-ph/0201018}}.

\bibitem{Donagi:2004su}
R.~Donagi, Y.-H. He, B.~A. Ovrut, and R.~Reinbacher, {\it Higgs doublets, split
  multiplets and heterotic su(3)c x su(2)l x u(1)y spectra},  {\em Phys. Lett.}
  {\bf B618} (2005) 259--264,
  [\href{http://xxx.lanl.gov/abs/hep-th/0409291}{{\tt hep-th/0409291}}].

\bibitem{Nath:2006ut}
P.~Nath and P.~F. Perez, {\it Proton stability in grand unified theories, in
  strings, and in branes},  \href{http://xxx.lanl.gov/abs/hep-ph/0601023}{{\tt
  hep-ph/0601023}}.

\bibitem{Tatar:2006dc}
R.~Tatar and T.~Watari, {\it Proton decay, yukawa couplings and underlying
  gauge symmetry in string theory},
  \href{http://xxx.lanl.gov/abs/hep-th/0602238}{{\tt hep-th/0602238}}.

\bibitem{Braun:2005xp}
V.~Braun, Y.-H. He, B.~A. Ovrut, and T.~Pantev, {\it Moduli dependent mu-terms
  in a heterotic standard model},
  \href{http://xxx.lanl.gov/abs/hep-th/0510142}{{\tt hep-th/0510142}}.

\bibitem{Braun:2006me}
V.~Braun, Y.-H. He, and B.~A. Ovrut, {\it Yukawa couplings in heterotic
  standard models},  \href{http://xxx.lanl.gov/abs/hep-th/0601204}{{\tt
  hep-th/0601204}}.

\bibitem{Giudice:1988yz}
G.~F. Giudice and A.~Masiero, {\it A natural solution to the mu problem in
  supergravity theories},  {\em Phys. Lett.} {\bf B206} (1988) 480--484.

\bibitem{Weinberg:2000cr}
S.~Weinberg, {\it The quantum theory of fields. vol. 3: Supersymmetry}, .
  Cambridge, UK: Univ. Pr. (2000) 419 p.

\bibitem{Lukas:1997rb}
A.~Lukas, B.~A. Ovrut, and D.~Waldram, {\it Gaugino condensation in m-theory on
  s**1/z(2)},  {\em Phys. Rev.} {\bf D57} (1998) 7529--7538,
  [\href{http://xxx.lanl.gov/abs/hep-th/9711197}{{\tt hep-th/9711197}}].

\bibitem{Riess:1998cb}
{\bf Supernova Search Team} Collaboration, A.~G. Riess {\em et.~al.}, {\it
  Observational evidence from supernovae for an accelerating universe and a
  cosmological constant},  {\em Astron. J.} {\bf 116} (1998) 1009--1038,
  [\href{http://xxx.lanl.gov/abs/astro-ph/9805201}{{\tt astro-ph/9805201}}].

\bibitem{Kachru:2003aw}
S.~Kachru, R.~Kallosh, A.~Linde, and S.~P. Trivedi, {\it De sitter vacua in
  string theory},  {\em Phys. Rev.} {\bf D68} (2003) 046005,
  [\href{http://xxx.lanl.gov/abs/hep-th/0301240}{{\tt hep-th/0301240}}].

\bibitem{Kachru:2002gs}
S.~Kachru, J.~Pearson, and H.~L. Verlinde, {\it Brane/flux annihilation and the
  string dual of a non- supersymmetric field theory},  {\em JHEP} {\bf 06}
  (2002) 021, [\href{http://xxx.lanl.gov/abs/hep-th/0112197}{{\tt
  hep-th/0112197}}].

\bibitem{Buchbinder:2003pi}
E.~I. Buchbinder and B.~A. Ovrut, {\it Vacuum stability in heterotic m-theory},
   {\em Phys. Rev.} {\bf D69} (2004) 086010,
  [\href{http://xxx.lanl.gov/abs/hep-th/0310112}{{\tt hep-th/0310112}}].

\bibitem{Buchbinder:2004im}
E.~I. Buchbinder, {\it Raising anti de sitter vacua to de sitter vacua in
  heterotic m-theory},  {\em Phys. Rev.} {\bf D70} (2004) 066008,
  [\href{http://xxx.lanl.gov/abs/hep-th/0406101}{{\tt hep-th/0406101}}].

\bibitem{Donagi:2000zf}
R.~Donagi, B.~A. Ovrut, T.~Pantev, and D.~Waldram, {\it Standard-model bundles
  on non-simply connected calabi-yau threefolds},  {\em JHEP} {\bf 08} (2001)
  053, [\href{http://xxx.lanl.gov/abs/hep-th/0008008}{{\tt hep-th/0008008}}].

\bibitem{Donagi:2004ia}
R.~Donagi, Y.-H. He, B.~A. Ovrut, and R.~Reinbacher, {\it The particle spectrum
  of heterotic compactifications},  {\em JHEP} {\bf 12} (2004) 054,
  [\href{http://xxx.lanl.gov/abs/hep-th/0405014}{{\tt hep-th/0405014}}].

\bibitem{Donagi:2004qk}
R.~Donagi, Y.-H. He, B.~A. Ovrut, and R.~Reinbacher, {\it Moduli dependent
  spectra of heterotic compactifications},  {\em Phys. Lett.} {\bf B598} (2004)
  279--284, [\href{http://xxx.lanl.gov/abs/hep-th/0403291}{{\tt
  hep-th/0403291}}].

\bibitem{Donagi:2004ub}
R.~Donagi, Y.-H. He, B.~A. Ovrut, and R.~Reinbacher, {\it The spectra of
  heterotic standard model vacua},  {\em JHEP} {\bf 06} (2005) 070,
  [\href{http://xxx.lanl.gov/abs/hep-th/0411156}{{\tt hep-th/0411156}}].

\bibitem{Diaconescu:2005pc}
D.-E. Diaconescu, B.~Florea, S.~Kachru, and P.~Svrcek, {\it Gauge - mediated
  supersymmetry breaking in string compactifications},  {\em JHEP} {\bf 02}
  (2006) 020, [\href{http://xxx.lanl.gov/abs/hep-th/0512170}{{\tt
  hep-th/0512170}}].

\bibitem{Derendinger:2000gy}
J.-P. Derendinger and R.~Sauser, {\it A five-brane modulus in the effective n =
  1 supergravity of m-theory},  {\em Nucl. Phys.} {\bf B598} (2001) 87--114,
  [\href{http://xxx.lanl.gov/abs/hep-th/0009054}{{\tt hep-th/0009054}}].

\bibitem{Lima:2001jc}
E.~Lima, B.~A. Ovrut, J.~Park, and R.~Reinbacher, {\it Non-perturbative
  superpotential from membrane instantons in heterotic m-theory},  {\em Nucl.
  Phys.} {\bf B614} (2001) 117--170,
  [\href{http://xxx.lanl.gov/abs/hep-th/0101049}{{\tt hep-th/0101049}}].

\bibitem{Lima:2001nh}
E.~Lima, B.~A. Ovrut, and J.~Park, {\it Five-brane superpotentials in heterotic
  m-theory},  {\em Nucl. Phys.} {\bf B626} (2002) 113--164,
  [\href{http://xxx.lanl.gov/abs/hep-th/0102046}{{\tt hep-th/0102046}}].

\bibitem{Lukas:1997fg}
A.~Lukas, B.~A. Ovrut, and D.~Waldram, {\it On the four-dimensional effective
  action of strongly coupled heterotic string theory},  {\em Nucl. Phys.} {\bf
  B532} (1998) 43--82, [\href{http://xxx.lanl.gov/abs/hep-th/9710208}{{\tt
  hep-th/9710208}}].

\bibitem{Candelas:1990pi}
P.~Candelas and X.~de~la Ossa, {\it Moduli space of calabi-yau manifolds},
  {\em Nucl. Phys.} {\bf B355} (1991) 455--481.

\bibitem{Lukas:1998hk}
A.~Lukas, B.~A. Ovrut, and D.~Waldram, {\it Non-standard embedding and
  five-branes in heterotic m- theory},  {\em Phys. Rev.} {\bf D59} (1999)
  106005, [\href{http://xxx.lanl.gov/abs/hep-th/9808101}{{\tt
  hep-th/9808101}}].

\bibitem{Buchbinder:2002wz}
E.~Buchbinder and B.~A. Ovrut, {\it Vector bundle moduli},  {\em Russ. Phys.
  J.} {\bf 45} (2002) 662--669.

\bibitem{Behrndt:2000zh}
K.~Behrndt and S.~Gukov, {\it Domain walls and superpotentials from m theory on
  calabi- yau three-folds},  {\em Nucl. Phys.} {\bf B580} (2000) 225--242,
  [\href{http://xxx.lanl.gov/abs/hep-th/0001082}{{\tt hep-th/0001082}}].

\bibitem{Becker:2002jj}
M.~Becker and D.~Constantin, {\it A note on flux induced superpotentials in
  string theory},  {\em JHEP} {\bf 08} (2003) 015,
  [\href{http://xxx.lanl.gov/abs/hep-th/0210131}{{\tt hep-th/0210131}}].

\bibitem{Dine:1985rz}
M.~Dine, R.~Rohm, N.~Seiberg, and E.~Witten, {\it Gluino condensation in
  superstring models},  {\em Phys. Lett.} {\bf B156} (1985) 55.

\bibitem{Kaplunovsky:1993rd}
V.~S. Kaplunovsky and J.~Louis, {\it Model independent analysis of soft terms
  in effective supergravity and in string theory},  {\em Phys. Lett.} {\bf
  B306} (1993) 269--275, [\href{http://xxx.lanl.gov/abs/hep-th/9303040}{{\tt
  hep-th/9303040}}].

\bibitem{Brignole:1993dj}
A.~Brignole, L.~E. Ibanez, and C.~Munoz, {\it Towards a theory of soft terms
  for the supersymmetric standard model},  {\em Nucl. Phys.} {\bf B422} (1994)
  125--171, [\href{http://xxx.lanl.gov/abs/hep-ph/9308271}{{\tt
  hep-ph/9308271}}].

\bibitem{Nilles:1997cm}
H.~P. Nilles, M.~Olechowski, and M.~Yamaguchi, {\it Supersymmetry breaking and
  soft terms in m-theory},  {\em Phys. Lett.} {\bf B415} (1997) 24--30,
  [\href{http://xxx.lanl.gov/abs/hep-th/9707143}{{\tt hep-th/9707143}}].

\bibitem{Horava:1996vs}
P.~Horava, {\it Gluino condensation in strongly coupled heterotic string
  theory},  {\em Phys. Rev.} {\bf D54} (1996) 7561--7569,
  [\href{http://xxx.lanl.gov/abs/hep-th/9608019}{{\tt hep-th/9608019}}].

\bibitem{Lalak:1997zu}
Z.~Lalak and S.~Thomas, {\it Gaugino condensation, moduli potential and
  supersymmetry breaking in m-theory models},  {\em Nucl. Phys.} {\bf B515}
  (1998) 55--72, [\href{http://xxx.lanl.gov/abs/hep-th/9707223}{{\tt
  hep-th/9707223}}].

\bibitem{Lukas:1999kt}
A.~Lukas, B.~A. Ovrut, and D.~Waldram, {\it Five-branes and supersymmetry
  breaking in m-theory},  {\em JHEP} {\bf 04} (1999) 009,
  [\href{http://xxx.lanl.gov/abs/hep-th/9901017}{{\tt hep-th/9901017}}].

\bibitem{Witten:1999eg}
E.~Witten, {\it World-sheet corrections via d-instantons},  {\em JHEP} {\bf 02}
  (2000) 030, [\href{http://xxx.lanl.gov/abs/hep-th/9907041}{{\tt
  hep-th/9907041}}].

\bibitem{Buchbinder:2002ic}
E.~I. Buchbinder, R.~Donagi, and B.~A. Ovrut, {\it Superpotentials for vector
  bundle moduli},  {\em Nucl. Phys.} {\bf B653} (2003) 400--420,
  [\href{http://xxx.lanl.gov/abs/hep-th/0205190}{{\tt hep-th/0205190}}].

\bibitem{Buchbinder:2002pr}
E.~I. Buchbinder, R.~Donagi, and B.~A. Ovrut, {\it Vector bundle moduli
  superpotentials in heterotic superstrings and m-theory},  {\em JHEP} {\bf 07}
  (2002) 066, [\href{http://xxx.lanl.gov/abs/hep-th/0206203}{{\tt
  hep-th/0206203}}].

\bibitem{Buchbinder:2002ji}
E.~Buchbinder, R.~Donagi, and B.~A. Ovrut, {\it Vector bundle moduli and small
  instanton transitions},  {\em JHEP} {\bf 06} (2002) 054,
  [\href{http://xxx.lanl.gov/abs/hep-th/0202084}{{\tt hep-th/0202084}}].

\bibitem{Beasley:2003fx}
C.~Beasley and E.~Witten, {\it Residues and world-sheet instantons},  {\em
  JHEP} {\bf 10} (2003) 065,
  [\href{http://xxx.lanl.gov/abs/hep-th/0304115}{{\tt hep-th/0304115}}].

\bibitem{toappear}
 to appear.

\bibitem{Dasgupta:1999ss}
K.~Dasgupta, G.~Rajesh, and S.~Sethi, {\it M theory, orientifolds and g-flux},
  {\em JHEP} {\bf 08} (1999) 023,
  [\href{http://xxx.lanl.gov/abs/hep-th/9908088}{{\tt hep-th/9908088}}].

\bibitem{Giddings:2001yu}
S.~B. Giddings, S.~Kachru, and J.~Polchinski, {\it Hierarchies from fluxes in
  string compactifications},  {\em Phys. Rev.} {\bf D66} (2002) 106006,
  [\href{http://xxx.lanl.gov/abs/hep-th/0105097}{{\tt hep-th/0105097}}].

\bibitem{Kachru:2002he}
S.~Kachru, M.~B. Schulz, and S.~Trivedi, {\it Moduli stabilization from fluxes
  in a simple iib orientifold},  {\em JHEP} {\bf 10} (2003) 007,
  [\href{http://xxx.lanl.gov/abs/hep-th/0201028}{{\tt hep-th/0201028}}].

\bibitem{Kachru:2002sk}
S.~Kachru, M.~B. Schulz, P.~K. Tripathy, and S.~P. Trivedi, {\it New
  supersymmetric string compactifications},  {\em JHEP} {\bf 03} (2003) 061,
  [\href{http://xxx.lanl.gov/abs/hep-th/0211182}{{\tt hep-th/0211182}}].

\bibitem{Frey:2002hf}
A.~R. Frey and J.~Polchinski, {\it N = 3 warped compactifications},  {\em Phys.
  Rev.} {\bf D65} (2002) 126009,
  [\href{http://xxx.lanl.gov/abs/hep-th/0201029}{{\tt hep-th/0201029}}].

\bibitem{Gukov:1999ya}
S.~Gukov, C.~Vafa, and E.~Witten, {\it Cft's from calabi-yau four-folds},  {\em
  Nucl. Phys.} {\bf B584} (2000) 69--108,
  [\href{http://xxx.lanl.gov/abs/hep-th/9906070}{{\tt hep-th/9906070}}].

\bibitem{Taylor:1999ii}
T.~R. Taylor and C.~Vafa, {\it Rr flux on calabi-yau and partial supersymmetry
  breaking},  {\em Phys. Lett.} {\bf B474} (2000) 130--137,
  [\href{http://xxx.lanl.gov/abs/hep-th/9912152}{{\tt hep-th/9912152}}].

\bibitem{Curio:2000sc}
G.~Curio, A.~Klemm, D.~Lust, and S.~Theisen, {\it On the vacuum structure of
  type ii string compactifications on calabi-yau spaces with h-fluxes},  {\em
  Nucl. Phys.} {\bf B609} (2001) 3--45,
  [\href{http://xxx.lanl.gov/abs/hep-th/0012213}{{\tt hep-th/0012213}}].

\bibitem{Curio:2003ur}
G.~Curio and A.~Krause, {\it Enlarging the parameter space of heterotic
  m-theory flux compactifications to phenomenological viability},  {\em Nucl.
  Phys.} {\bf B693} (2004) 195--222,
  [\href{http://xxx.lanl.gov/abs/hep-th/0308202}{{\tt hep-th/0308202}}].

\bibitem{Balasubramanian:2004uy}
V.~Balasubramanian and P.~Berglund, {\it Stringy corrections to kaehler
  potentials, susy breaking, and the cosmological constant problem},  {\em
  JHEP} {\bf 11} (2004) 085,
  [\href{http://xxx.lanl.gov/abs/hep-th/0408054}{{\tt hep-th/0408054}}].

\bibitem{Balasubramanian:2005zx}
V.~Balasubramanian, P.~Berglund, J.~P. Conlon, and F.~Quevedo, {\it Systematics
  of moduli stabilisation in calabi-yau flux compactifications},  {\em JHEP}
  {\bf 03} (2005) 007, [\href{http://xxx.lanl.gov/abs/hep-th/0502058}{{\tt
  hep-th/0502058}}].

\bibitem{Weinberg:1987dv}
S.~Weinberg, {\it Anthropic bound on the cosmological constant},  {\em Phys.
  Rev. Lett.} {\bf 59} (1987) 2607.

\bibitem{Choi:2004sx}
K.~Choi, A.~Falkowski, H.~P. Nilles, M.~Olechowski, and S.~Pokorski, {\it
  Stability of flux compactifications and the pattern of supersymmetry
  breaking},  {\em JHEP} {\bf 11} (2004) 076,
  [\href{http://xxx.lanl.gov/abs/hep-th/0411066}{{\tt hep-th/0411066}}].

\bibitem{Choi:2005ge}
K.~Choi, A.~Falkowski, H.~P. Nilles, and M.~Olechowski, {\it Soft supersymmetry
  breaking in kklt flux compactification},  {\em Nucl. Phys.} {\bf B718} (2005)
  113--133, [\href{http://xxx.lanl.gov/abs/hep-th/0503216}{{\tt
  hep-th/0503216}}].

\end{thebibliography}\endgroup

\end{document}